\documentclass[twocolumn,showpacs,aps,prd,amssymb,nofootinbib,nobibnotes,floatfix]{aastex631}
\usepackage{hyperref}
\usepackage{amsfonts}
\usepackage{amsmath}
\usepackage{mathrsfs}
\usepackage{epsfig}
\usepackage{graphicx}
\usepackage{url}
\usepackage{hyperref}
\usepackage{float}
\usepackage{color}
\usepackage[utf8]{inputenc} 
\usepackage{epsf}
\usepackage{comment}
\usepackage{slashed}
\usepackage{footmisc}
\definecolor{linkcolor}{rgb}{0.0,0.3,0.5}
\usepackage[caption=false]{subfig}
\hypersetup{linkcolor=magenta,citecolor=cerulean,filecolor=cyan,urlcolor=magenta}

\def\PN{{\cal PN}}

\usepackage{color}
\definecolor{cerulean}{rgb}{0.0, 0.48, 0.65}

\definecolor{navy}{rgb}{0.2, 0.0, 1.0}

\definecolor{jungle}{rgb}{0.0, 0.5, 0.0}

\usepackage{color}

\begin{document}

\title{Gravitational Wave Phase Shifts in Eccentric Black Hole Mergers\\
as a Probe of Dynamical Formation Environments}

\author{Johan Samsing$^{1}$, Kai Hendriks$^{1}$, Lorenz Zwick$^{1}$, Daniel J. D'Orazio$^{1}$, Bin Liu}
\affiliation{\vspace{2mm}Niels Bohr International Academy, The Niels Bohr Institute, Blegdamsvej 17, DK-2100, Copenhagen, Denmark}
\affiliation{\vspace{2mm}Institute for Astronomy, School of Physics, Zhejiang University, 310058 Hangzhou, China.}

\correspondingauthor{Johan Samsing}
\email{jsamsing@gmail.com}

\begin{abstract}

We quantify for the first time the gravitational wave (GW) phase shift appearing in the waveform of eccentric binary black hole (BBH)
mergers formed dynamically in three-body systems. For this, we have developed a novel numerical method where we construct a reference binary, by evolving
the post-Newtonian ($\PN$) evolution equations backwards from a point near merger without the inclusion of the third object, that can be compared to the
real binary that evolves under the influence from the third BH. From this we quantify how the interplay between dynamical tides, $\PN$-effects, and the time-dependent
Doppler shift of the eccentric GW source results in unique observable GW phase shifts that can be mapped to the gravitational dynamics taking place at formation.
We further find a new analytical expression for the GW phase shift, which surprisingly has a universal functional form
that only depends on the time-evolving BBH eccentricity. The normalization scales with the BH masses and initial separation,
which can be linked to the underlying astrophysical environment.
GW phase shifts from a chaotic 3-body BH scattering taking place in a cluster, and from a BBH inspiraling in a disk migration trap near a super-massive BH, are also shown for illustration.
When current and future GW detectors start to observe eccentric GW sources with high enough signal-to-noise-ratio, we propose
this to be among the only ways of directly probing the dynamical origin of individual BBH mergers using GWs alone.

\end{abstract}

\section{Introduction}\label{sec:Introduction}

The number of merging binary black holes (BBHs) observed in gravitational waves (GWs) with LIGO/Virgo/Kagra (LVK) is steadily increasing \citep{2023ApJS..267...29A}.
However, how these mergers form in our Universe is still a major unsolved question, and with the wide
variety of observed BH spins \citep{2019PhRvD.100b3007Z, 2021PDU....3100791G}, masses \citep{2019ApJ...882L..24A, 2020PhRvL.125j1102A}, and possible
eccentricity \citep{2019ApJ...883..149A, 2021ApJ...921L..31R, 2022NatAs...6..344G, 2023arXiv230803822T},
Nature likely assemble these BBHs in a variety of different ways.
Some proposed formation environments include dense stellar clusters \citep{2000ApJ...528L..17P, Lee:2010in,
2010MNRAS.402..371B, 2013MNRAS.435.1358T, 2014MNRAS.440.2714B,
2015PhRvL.115e1101R, 2015ApJ...802L..22R, 2016PhRvD..93h4029R, 2016ApJ...824L...8R,
2016ApJ...824L...8R, 2017MNRAS.464L..36A, 2017MNRAS.469.4665P, 2018PhRvD..97j3014S, 2018MNRAS.tmp.2223S, 2019arXiv190711231S, 2021MNRAS.504..910T, 2022MNRAS.511.1362T},
isolated binary stars \citep{2012ApJ...759...52D, 2013ApJ...779...72D, 2015ApJ...806..263D, 2016ApJ...819..108B,
2016Natur.534..512B, 2017ApJ...836...39S, 2017ApJ...845..173M, 2018ApJ...863....7R, 2018ApJ...862L...3S, 2023MNRAS.524..426I},
active galactic nuclei (AGN) discs \citep{2017ApJ...835..165B,  2017MNRAS.464..946S, 2017arXiv170207818M, 2020ApJ...898...25T, 2022Natur.603..237S, 2023arXiv231213281T, Fabj24},
galactic nuclei (GN) \citep{2009MNRAS.395.2127O, 2015MNRAS.448..754H,
2016ApJ...828...77V, 2016ApJ...831..187A, 2016MNRAS.460.3494S, 2017arXiv170609896H, 2018ApJ...865....2H,2019ApJ...885..135T, 2019ApJ...883L...7L,2021MNRAS.502.2049L, 2023MNRAS.523.4227A},
very massive stellar mergers \citep{Loeb:2016, Woosley:2016, Janiuk+2017, DOrazioLoeb:2018},
and single-single GW captures of primordial black holes \citep{2016PhRvL.116t1301B, 2016PhRvD..94h4013C,
2016PhRvL.117f1101S, 2016PhRvD..94h3504C}.
But how do we observationally disentangle these channels?

The success of LVK only marks the beginning of this new field,
as new proposed observatories, such as the earth-based 3G-observatories the Einstein Telescope (ET) \citep{2020JCAP...03..050M} and Cosmic Explorer (CE) \citep{2023arXiv230613745E},
and the space-borne DECIGO/TianQin/Taiji 
\citep{2011CQGra..28i4011K, 2016CQGra..33c5010L, 10.1093/nsr/nwx116, 2020PhRvD.101j3027L} (deci-Hertz) and LISA \citep{2017arXiv170200786A} (milli-Hertz),
not only will expand our observable window in GW frequency allowing for multi-band GW
observations \citep[e.g.][]{2016PhRvL.116w1102S}, but also will be much more sensitive leading to an estimated $\sim 10^{5}$ observed mergers per year near the current LVK bands.
This will not only make it more likely to observe some of the rare and more scientifically valuable GW sources,
but also make it possible to build up solid statistics in order to gain new insight into the compact-object merger history as a function of
redshift \citep[e.g.][]{2018ApJ...863L..41F, 2022ApJ...932L..19B}, as well as constraining cosmology \citep[e.g.][]{2019ApJ...883L..42F}.

The variety of proposed formation channels is large, and it seems very challenging with current methods to tell individual channels apart using GWs alone. However,
certain groups of channels are likely to be distinguishable from each other, e.g., BBHs forming through dynamical interactions will give rise to a
significant fraction of eccentric mergers \citep[e.g.][]{2006ApJ...640..156G, 2014ApJ...784...71S, 2017ApJ...840L..14S, Samsing18a, Samsing2018, Samsing18, 2018ApJ...855..124S,
2018MNRAS.tmp.2223S, 2018PhRvD..98l3005R, 2019ApJ...881...41L,2019ApJ...871...91Z, 2019PhRvD.100d3010S, 2019arXiv190711231S} with distinct distributions in
LVK \citep{Samsing18}, DECIGO/TianQin/Taiji \citep[e.g.][]{2017ApJ...842L...2C, 2019arXiv190711231S},
and LISA \citep{2018MNRAS.tmp.2223S,2018MNRAS.481.4775D, 2019PhRvD..99f3003K}, in contrast to those forming through isolated binary
evolution. Other parameters include the relative spin orientation of the
merging BBHs \citep[e.g.][]{2000ApJ...541..319K, 2016ApJ...832L...2R,2018ApJ...863...68L}, as well as the mass
spectrum \citep[e.g.][]{2017ApJ...846...82Z,2021MNRAS.505.3681S}.
Although this is encouraging, the group of dynamically formed BBH mergers consists likewise of several sub-channels that
again overlap in their observed quantities. For example, systems in hierarchical Lidov-Kozai-triple configurations \citep[e.g.][]{2018ApJ...856..140H, 2019ApJ...881...41L},
GW captures forming in GN \citep[e.g.][]{2009MNRAS.395.2127O}, and few-body interactions in globular cluster (GCs) \citep[e.g.][]{2018PhRvD..97j3014S} and
AGNs \citep[e.g.][]{2022Natur.603..237S, Fabj24} all give rise to eccentric mergers. The question in this case is therefore how do we tell the difference
between the proposed dynamical channels?

In this paper, we explore and quantify an observable that can be used to distinguish eccentric BBH mergers formed dynamically through
strong few-body interactions from those formed instead in near isolation, such as through single-single GW captures. The idea is that if a
third object is nearby, as in the case of a binary-single mediated merger,
the presence of this object will lead to modulations (dynamical and relativistic) in the observed GW-form that are not present, or cannot be trivially
mimicked by, e.g., changing the BH mass and spin, in the GW-form of an isolated merging BBH.
These modulations will, to leading order, manifest as a phase shift in the observed GWs emitted by the
eccentric binary, from which one in principle can infer the orbital evolution of all three interacting objects, in a similar way to what
is done in the binary pulsar case \citep[e.g.][]{2017ApJ...834..200M}. 
This represents one of the only ways to probe the assembly mechanism of individual GW sources using GWs alone.
In fact, a (circular) GW source observed by LVK might already show indications of GW phase shift \citep{2024arXiv240101743H}.
Similar ideas have been presented for LISA-like sources, where the GW evolution time is long enough for a change in center-of-mass
(COM) motion to be observed over the duration of the mission
\citep[e.g.][]{2011PhRvD..83d4030Y, 2017PhRvD..96f3014I, 2018PhRvD..98f4012R, 2019PhRvD..99b4025C, 2019ApJ...878...75R, 2019MNRAS.488.5665W,
2020PhRvD.101f3002T, 2020PhRvD.101h3031D, 2021PhRvL.126j1105T, 2022PhRvD.105l4048S, 2023PhRvD.107d3009X, 2023arXiv231016799L}.
This line of models have recently been extended to include higher order General Relativistic (GR) corrections and spin effects \citep[e.g.][]{2023PhRvD.107h4011C, 2023arXiv231006894C}.
A few papers have also explored the effect across several GW bands, including the
LVK bands \citep[e.g.][]{2023ApJ...954..105V, 2024MNRAS.527.8586T, 2023ApJ...954..105V}, but using simplified assumptions.
However, the general case of dynamically assembled eccentric sources, including those forming naturally during chaotic scatterings in GCs \citep[e.g.][]{2018PhRvD..97j3014S, 2018PhRvD..98l3005R},
have not yet been consistently explored using dynamical prescriptions as the one we introduce in this paper.
GW phase shift effects are not limited to dynamics alone, e.g., gas-drag and other dissipative effects will also give rise to their
own unique GW perturbations \citep[e.g.][]{2014barausse, 2023MNRAS.521.4645Z}.

With this motivation, we here present an exploration of how a third object perturbs and leads to shifts in the GW phase of a dynamically formed
eccentric GW source from assembly to merger for both idealized and chaotic cases.
This includes both an analytical solution, as well as a novel numerical post-Newtonian ($\PN$) framework that we use to directly quantify
the non-linear dynamical influence on the merging BBH from the perturber, and how this gives rise to unique GW phase shifts. 
Putting aside the observable challenges of eccentric sources, the basic questions we explore in this paper are
under what mass-, length-, and time-scales the GW phase shift is expected to be detectable, and if this could happen in real astrophysical systems.
We further point out that the leading GW phase shift effect can be
constrained using standard GW template matching using already existing template banks derived without these effects \citep[e.g.][]{2024arXiv240101743H}. The ongoing progress in
developing waveforms for eccentric sources therefore holds a tremendous potential to not only identify dynamically assembled mergers,
but also to learn about the exact few-body dynamics that brought the BBHs together.

\section{Setup and Numerical Methods}\label{sec:Methods and Numerical Setup}

Below we introduce our setup, numerical methods, and describe what physical effects
that generally result in GW phase shifts, as well as how this dependent on the location of the observer.

\subsection{Example and Setup}\label{sec:Example}

\begin{figure*}
\centering
\includegraphics[width=\textwidth]{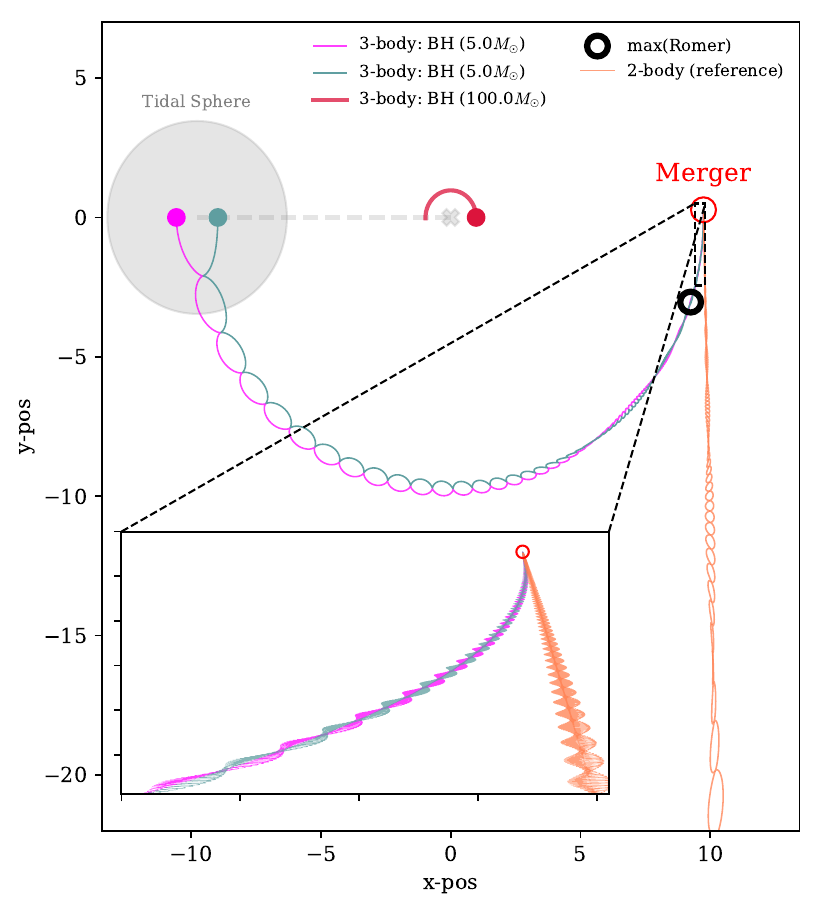}
\caption{{\bf Evolution of a bound eccentric binary black hole merger:} Evolution of a binary black hole (BH1, BH2) with masses $m_1 = m_2 = 5M_{\odot}$, orbiting
a single black hole BH3 with mass $m_3 = 100M_{\odot}$ on a circular outer orbit with radius $R = 0.05\ AU$. The binary initially starts within the Hill sphere shown to the left (grey),
after which it orbits BH3 for then to merge at the right at the point `Merger'. Near merger is attached a `2-body-reference binary' (orange),
which shows the evolution if the central BH3 is not present. This reference binary is generated by evolving the $\PN$-equations
backwards in time from the point near merger,
as further described in Sec. \ref{sec:Comparing to a 2-body-reference Binary}. We quantify the observable
effects from the presence of BH3, by comparing the real `3-body-perturbed binary' to this `2-body-reference binary', as further explained in
Sec. \Ref{sec:Example and Numerical Methods}.}
\label{fig:Ex1_orbits}
\end{figure*}

In this paper we focus on BBH GW sources that form dynamically in three-body systems, and to what degree the presence of the third
object is able to modulate the observable GW signal emitted from the eccentric inspiraling BBH.
To introduce the problem and our numerical methods, we first
consider the three-body setup shown in Fig. \ref{fig:Ex1_orbits}. This shows a `binary' BH consisting of BH1($m_1 = 5M_{\odot}$)
and BH2($m_2 = 5M_{\odot}$) with initial semi-major axis (SMA), $a_0$, and eccentricity, $e_0$, on an outer circular orbit with radius $R$ around
a `perturber' denoted BH3($m_3 = 100 M_{\odot}$). In general, throughout the paper, we will denote the binary components `1' and `2', and the perturber by `3',
often with a `BH' in front. The (BH1,BH2)-binary in Fig. \ref{fig:Ex1_orbits} is initiated within the binary Hill radius given by,
\begin{equation}
R_H \sim R((m_1+m_2)/m_3)^{1/3}.
\label{eq:RH}
\end{equation}
From this initial position, it orbits counter clockwise to finally merge,
through the emission of GWs, after about half an outer orbit.
In the following we describe how we evolve such systems,
and how to numerically extract the GW phase shift.

\subsection{Numerical Methods}\label{sec:Example and Numerical Methods}

The numerical methods we use here and throughout the paper are based on the $\PN$-formalism \citep{Blanchet06, Blanchet14}, where the $N$-body
equations-of-motion (EOM) can be written in a `Newtonian way', where the inclusion of GR effects are added to the Newtonian acceleration
through a series-expansion in $(1/c)$, where $c$ is the speed of light. Following the notation by \cite{2006MNRAS.371L..45K}, the acceleration equation with the inclusion of $\PN$ terms
can be written as,
\begin{equation}
\underline{a}  = \underbrace{\underline{a}_0}_{\rm Newt.}+
\underbrace{\underbrace{c^{-2}\underline{a}_2}_{1\PN}\ +
\underbrace{c^{-4}\underline{a}_4}_{2\PN}}_{\rm precession}\ +
\underbrace{\underbrace{c^{-5}\underline{a}_5}_{2.5\PN}}_{\rm GW~rad.} +\ 
\mathcal{O},
\label{eq:a_expansionPN}
\end{equation}
where ${\underline{a}_0}$ is the classical Newtonian acceleration term (Newt.),
\begin{equation}
\underline{a}_0 = \frac{Gm_2}{r^2}\underline{n},
\label{eq.a0}
\end{equation}
the following two terms, $\underline{a}_2$ and $\underline{a}_4$, are conservative terms leading to precession (1$\PN$, 2$\PN$),
\begin{align}
\underline{a}_2
& =\frac{Gm_2}{r^2} \Big( \underline{n}\Big[-v_1^2-2v_2^2+4v_1v_2+\frac{3}{2}(nv_2)^2+\nonumber\\
& 5\frac{Gm_1}{r}+4\frac{Gm_2}{r}\Big]+(\underline{v}_1-\underline{v}_2)
\big[4nv_1-3nv_2\big] \Big),
\label{eq.a2}
\end{align}
\begin{align}
\underline{a}_4
& = \frac{Gm_2}{r^2}\mbox{\Big(}\underline{n}
\mbox{\Big[}-2v_2^4+4v_2^2(v_1v_2)-2(v_1v_2)^2\nonumber\\
&+\frac{3}{2}v_1^2(nv_2)^2+\frac{9}{2}v_2^2(nv_2)^2-6(v_1v_2)(nv_2)^2\nonumber\\
&-\frac{15}{8}(nv_2)^4+\frac{Gm_1}{r}\mbox{\Big(}-\frac{15}{4}v_1^2+
\frac{5}{4}v_2^2-\frac{5}{2}v_1v_2\nonumber\\
&+\frac{39}{2}(nv_1)^2-39(nv_1)(nv_2)+\frac{17}{2}(nv_2)^2\mbox{\Big)}\nonumber\\
&+\frac{Gm_2}{r}(4v_2^2-8v_1v_2+2(nv_1)^2\nonumber\\
&-4(nv_1)(nv_2)-6(nv_2)^2)\mbox{\Big]}\nonumber\\
&+(\underline{v}_1-\underline{v}_2)\mbox{\Big[}v^2_1(nv_2)+4v_2^2(nv_1)-5v_2^2(nv_2)\nonumber\\
&-4(v_1v_2)(nv_1)+4(v_1v_2)(nv_2)-6(nv_1)(nv_2)^2\nonumber\\
&+\frac{9}{2}(nv_2)^3+\frac{Gm_1}{r}\big(-\frac{63}{4}nv_1+\frac{55}{4}nv_2\big)\nonumber\\
&+\frac{Gm_2}{r} \big(-2nv_1-2nv_2\big)\mbox{\Big]}\mbox{\Big)}\nonumber\\
&+\frac{G^3m_2}{r^4} \mbox{\Big(} \underline{n}\big[-\frac{57}{4}m_1^2-9m_2^2-\frac{69}{2}m_1m_2\big]\mbox{\Big)},
\label{eq.a4}
\end{align}
and the last term, $\underline{a}_5$, is the leading term in the series that is dissipative (2.5$\PN$), which results in pairwise
energy and angular momentum loss through the emission of GWs,
\begin{align}
\underline{a}_5   
& =\frac{4}{5}\frac{G^2m_1m_2}{r^3}\mbox{\Big(} (\underline{v}_1-
\underline{v}_2)\Big[-(\underline{v}_1-\underline{v}_2)^2\nonumber\\
& +2\frac{Gm_1}{r} -8\frac{Gm_2}{r}\Big]\nonumber\\
&+\underline{n}(nv_1-nv_2)\big[3(\underline{v}_1-\underline{v}_2)^2-6\frac{Gm_1}{r}
+\frac{52}{3}\frac{Gm_2}{r}\big]\mbox{\Big)}.
\label{eq.a5}
\end{align}
In these equations, $\underline{a}$ denotes the acceleration vector of object `1' from object `2',
$m_i$ and $\underline{v}_i$ denotes the mass and velocity vector of object `i', respectively, $r$ is the scalar distance from object `2' to `1',
$\underline{n}$ is the corresponding unit vector pointing from `2' to `1', and $x_{1}x_2$ denotes the dot-product of the two vectors $\underline{x}_1$ and $\underline{x}_2$. Note here that this pairwise acceleration
must be evaluated in the center-of-mass (COM) of the two objects; the velocities are therefore relative to the pairwise COM.
In the following sections, we discuss how to quantify the modulations caused by the
the third object, the role of the different $\PN$ terms, and what the main resultant observable effects are.

\subsection{Comparing to a 2-body Reference Binary}\label{sec:Comparing to a 2-body-reference Binary}

To numerically quantify the effect from the nearby perturber, BH3, on the evolution and emission of GWs from the (BH1,BH2)-binary, we now introduce a
{\it 2-body Reference Binary} defined as a binary with the same properties near merger as the 3-body-perturbed binary, but with a backwards
trajectory from the time of merger that follows the evolution equations as if the perturbing BH3 was not present.
In other words, our 2-body-reference binary refers to a normal unperturbed isolated GW inspiral between BH1 and BH2 that has the same properties
right around merger as the 3-body-perturbed GW inspiral, but a different evolution as one goes backwards in time compared to the 3-body-perturbed
binary evolution. To be able to make a correct comparison between
our 3-body-perturbed binary and our 2-body-reference binary, we set the reference binary COM velocity vector to be the
same as that of the perturbed binary right before merger. From this we account for the trivial difference that could arise from the constant (longitudinal) Doppler shift,
which also is not measurable due to its direct degeneracy with mass and redshift.
Note that this way of quantifying differences by introducing a `reference binary' was also used in the theoretical arguments by \cite{2017ApJ...834..200M}. 

The `2-body-reference binary' is seen in Fig. \ref{fig:Ex1_orbits} with {\it orange lines}, and is calculated by simulating the
(BH1,BH2)-binary backwards in time without BH3 from a point right before merger, by inverting the `time' and flipping the `sign' of the 2.5$\PN$
radiation term. By now comparing the 3-body 
evolution of the eccentric BBH around BH3 (`3-body (perturbed)') with the 2-body-reference binary evolution (`2-body (reference)'), as e.g. seen in Fig. \ref{fig:Ex1_orbits},
we can resolve, for the first time, the GW phase shift arising from the
interplay between tidal couplings, $\PN$-terms, and the finite propagation speed of GWs giving rise to what is known as the
{\it Rømer delay}, as is described in much greater detail in Sec. \ref{sec:GW Time Delays Along the Orbit} and Fig. \ref{fig:BS_ill}.

\subsection{Modulations of the GW signal}\label{sec:GW Perturbation Effects}

Several effects can leave an imprint on the GW-form from the real perturbed inspiraling BBH, compared to the isolated reference binary, including
dynamics, Doppler effects, Shapiro- and GR time-delays, strong- and diffractive-lensing, as well as GW boosting (see \cite{2017ApJ...834..200M}
for a compilation and comparison of some of these effects); However, in this paper we restrict our self to first include and study the following two distinct effects:
\begin{itemize}
\item{{\bf Dynamics:} Using our $N$-body code and novel numerical setup, we follow the Newtonian- and post-Newtonian gravitational interactions between all objects,
and from this keep track of how these couplings influence the evolution of the inspiraling BBH from assembly to merger. This dynamically perturbed evolution will
show up as a GW phase shift.}
\item{{\bf Rømer time-delay:} Since GWs propagate at the speed of light, the arrival time in the observer frame will depend on the orbital trajectory of the
BBH source as it inspirals, which will give rise to a time-dependent GW phase shift. Using our $\PN$-body code, we derive this shift by comparing a 2-body-reference binary
with our 3-body-perturbed binary.}
\end{itemize}
The other effects listed above will occasionally also play a role, which depends not only on the system, but also on the location of the observer. For example,
the GR time-delay could play a role if the outer orbit has a significant eccentricity, as this time-delay essentially
is a mapping of the potential difference the source experiences along its orbit. In addition, the observed delay from GR will to leading order not depend on the position
of the observer, and could therefore dominate over e.g. Doppler effects that clearly are largest when observing in the same plane as the source is moving in.

\begin{figure}
\centering
\includegraphics[width=\columnwidth]{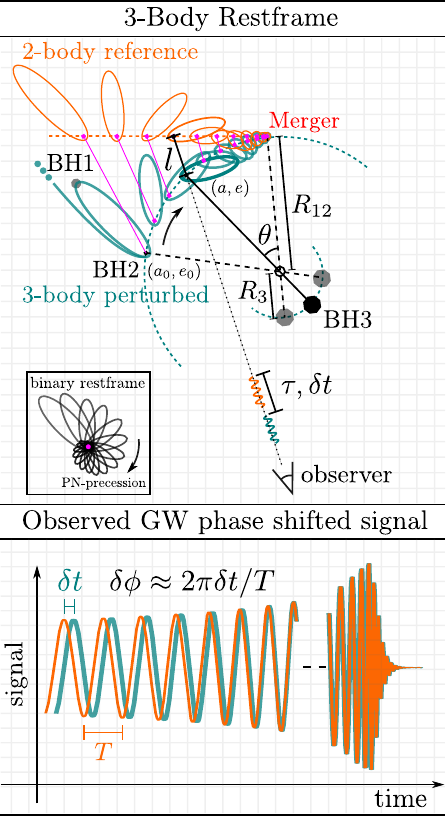}
\caption{{\bf Time delay and GW phase shifts:}
{\it Top plot:} Illustration of a (BH1,BH2)-binary, `3-body-perturbed', with initial elements $a_0, e_0$, orbiting
a third object, BH3, until merger takes place at `Merger'.
At `Merger' is attached a reference binary (2-body-reference) representing the trajectory the
(BH1,BH2)-binary would have taken if BH3 was not present (see Sec. \ref{sec:Comparing to a 2-body-reference Binary}).
Several effects can course the observed GW signal from the two evolutionary paths to
differ (see Sec. \ref{sec:GW Phase Shifts}),
such as the effect from traveling on a curved- vs. a straight path, as shown here.
The two paths differ by a maximum distance $l$ (pink lines) that gives rise to a
maximum time shift $\tau = l/c$, which manifests as an observed value $\delta{t}$.
{\it Bottom plot:} Illustration of the 3-body-perturbed and 2-body
reference GW signals, respectively. The two signals are displaced
relative to each other, which is what is referred to as the GW phase
shift. This shift can be
approximated by $\delta{\phi}(t) \approx 2\pi \delta{t}(t)/T(t)$, where $T(t)$
is the orbital period of the reference binary, as further described in Sec. \ref{sec:GW Phase Shifts}.
}
\label{fig:BS_ill}
\end{figure}

In the limit where the inspiraling BBH is only moving a small fraction of its orbit around its perturber, the total GW phase shift as
seen by a distant stationary observer, $\delta{\phi}$, can be approximated as \citep[e.g.][]{2017ApJ...834..200M},
\begin{align}
\delta{\phi} & \approx \delta{\phi}_{RDL} + \delta{\phi}_{DYN} + \delta{\phi}_{GR} \nonumber\\
\delta{\phi} & \approx \Delta{\phi}_{RDL} \times cos(i)sin(j) + \delta{\phi}_{DYN} + \delta{\phi}_{GR},
\label{eq:ADD_GWphs}
\end{align}
where $\delta{\phi}_{RDL}$ is the GW phase shift from Rømer Delay (RDL) as inferred by a distant observer,
$\Delta{\phi}_{RDL}$ is the {\it maximum possible} GW phase shift from Rømer Delay (see Fig. \ref{fig:BS_ill}),
$i$ is the angle in the plane of the `outer binary' (the binary composed of (BH1,BH2)+BH3) between the line-of-sight (LOS) of the
observer and the line connecting the location of BBH merger and the COM of the 3-body system, $j$ is the angle between the LOS and the angular momentum vector of the `outer binary',
$\delta{\phi}_{DYN}$ is the contribution from dynamics (DYN), and $\delta{\phi}_{GR}$ refers to the remaining
GR-effects (Shapiro delay, GR time-delay, etc.). In this work we study the contributions
$\delta{\phi} \approx \Delta{\phi}_{RDL} \times cos(i)sin(j) + \delta{\phi}_{DYN}$, and will in the remaining parts of the paper
simply refer to the total GW phase shift seen by an observer by $\delta{\phi}$, and the maximum GW phase shift component from Rømer delay by $\Delta{\phi}$.

\subsection{Observers and GW Propagation}\label{sec:Numerical Setup and Observers}

For studying and quantifying the GW phase shift induced by
dynamics and Rømer delays (finite propagation speed of GWs), we consider an observer at a fixed location far away from the
system:
\\
\\
{\it In all examples shown in this paper, we place the observer in the plane of the 3-body system along a line connecting the
3-body COM and the point of merger. We refer to this as the `Observer'.}
\\
\\
In terms of effects related to Rømer delays, this location will result in
GW phase shifts that are close to the maximum possible value, i.e. $\delta{\phi}_{RDL}\approx \Delta{\phi}_{RDL}$, as we will show and describe later.

Numerically, for every spatial ($x,y,z$)-point evaluated by our $N$-body code along each binary trajectory (`3-body-perturbed' and `2-body-reference')
we propagate a `GW signal' from that point to the observer by taking into account that the GWs travel at the speed of light, $c$. Again, for this
we do not include possible GR wave scattering effects caused by the perturber \citep[e.g.][]{2018PhRvD..98j4029D}, i.e. we assume the signal travels on straight
lines between the source and the observer. This setup is also sketched out in Fig. \ref{fig:BS_ill}. In this way we can follow and compare different
quantities of the eccentric inspiraling BBH as a function of time in the rest-frame of the observer, as will be shown below.

\section{Numerical Results}\label{sec:Gravitational Wave Phase Shifts}

In the following sections we quantify the observable imprints on the binary GW signal from the presence of the BH perturber for
our example shown in Fig. \ref{fig:Ex1_orbits}, by systematically studying the difference between different quantities of the
3-body-perturbed binary and the 2-body-reference binary. This will also serve as a general explanation for how a perturber
will give rise to a GW phase shift.

\subsection{Tidal Perturbations of Binary Elements}\label{sec:Tidal perturbations of Binary Elements}

\begin{figure}
\centering
\includegraphics[width=\columnwidth]{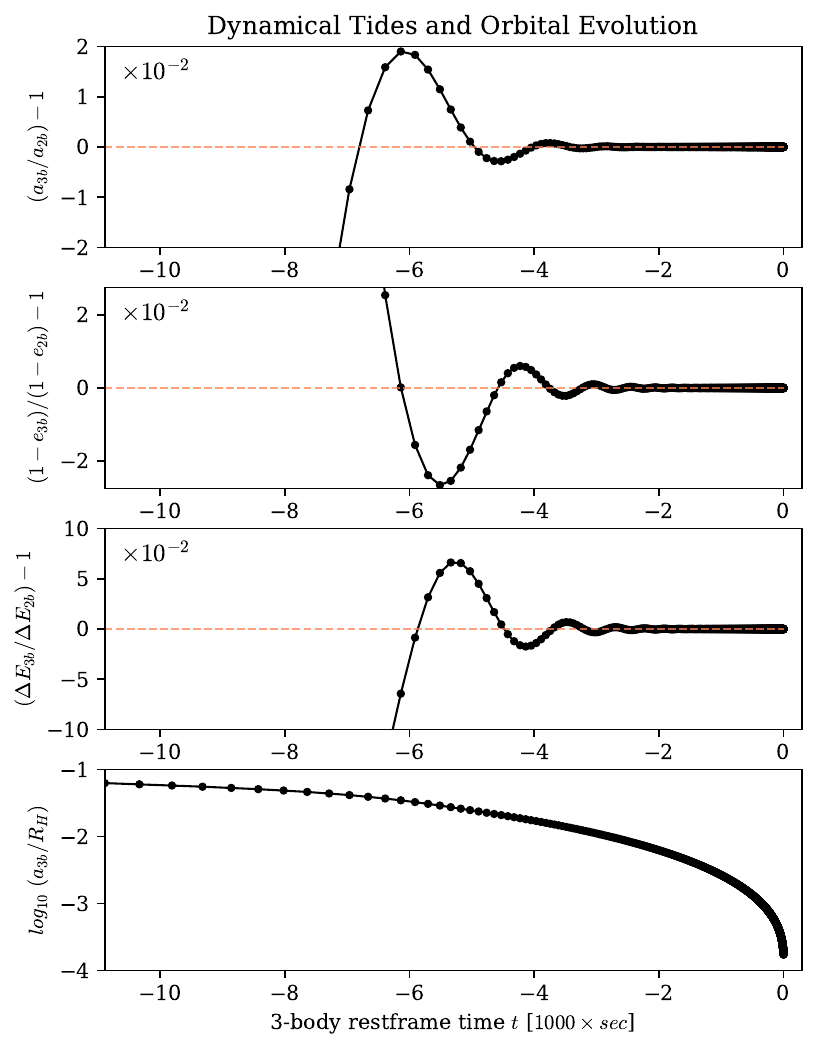}
\caption{{\bf Change in orbital elements from tides:}
The figures shows data from the illustrative setup presented in Fig. \ref{fig:Ex1_orbits}, of a
$[5M_{\odot}, 5M_{\odot}]$ BBH orbiting a $100 M_{\odot}$ BH perturber.
Counting from top to bottom:
{\it Panel 1:} Relative difference in SMA between the 3-body-perturbed binary ($a_{3b}$) and the 2-body-reference binary ($a_{2b}$),
as a function of time.
{\it Panel 2:} Relative difference in eccentricity as a function of time.
{\it Panel 3:} Corresponding relative difference in GW energy radiated at pericenter, $\Delta{E}_{3b}/\Delta{E}_{2b} - 1 = (r_{3b}/r_{2b})^{-7/2} - 1$
(see Eq. \ref{eq:DE_GW}), where $r = a(1-e)$ is the pericenter distance.
{\it Panel 4:} SMA of the 3-body-perturbed BBH relative to the Hill sphere (see Eq. \ref{eq:RH}).
All orbital quantities have here been estimated using the geometric method, as further explained in
Sec. \ref{sec:Tidal perturbations of Binary Elements} and shown in Eq. \ref{eq:ae_geom}.}
\label{fig:tides_orbitele}
\end{figure}

The first quantities we study are the SMA and eccentricity of the inspiraling BBH, and how the tidal influence of BH3 affect these orbital elements
as a function time in the rest-frame of the binary, i.e. without any propagation effects including Rømer delays.
For this we now consider Fig. \ref{fig:tides_orbitele}, which shows from top to bottom the difference between the orbital elements of the 3-body-perturbed binary and
the 2-body-reference binary in terms of the relative change in semi-major axis ($a_{3b}/a_{2b} - 1$),
eccentricity ($(1-e_{3b})/(1-e_{2b})-1$), GW energy radiated at pericenter ($\Delta{E_{3b}}/\Delta{E_{2b}} - 1$),
and the semi-major axis relative to the binary Hill-sphere, ($a_{3b}/R_{H}$), for the system presented in Fig \ref{fig:Ex1_orbits}.
Here the subscripts $3b$ and $2b$ refer to the 3-body-perturbed binary and 2-body-reference binary, respectively.
The SMA and the eccentricity, $a$ and $e$, shown here are estimated using a geometric method that is based on knowing the position of
the interacting objects as a function of time, such that $a$ and $e$ can be estimated as,
\begin{align}
a	& = (r_a + r_p)/2 \nonumber\\
e	& = 1-r_p/a
\label{eq:ae_geom}
\end{align}
where $r_a$ and $r_p$ is the apocenter- and pericenter-distance of the time evolving BBH orbit, respectively,
that we measure using our $N$-body code. In fact, the Newtonian orbital elements, $a$ and $e$, can be very difficult to even proper
define for strong precessing and decaying binaries, and can as a result lead to biased results
\citep[e.g.][]{2020MNRAS.495.2321Z, 2021PhRvD.104j4023T, 2022PhRvD.106j4035C, 2023PhRvD.107f4024B, 2023PhRvD.108j4007S}.

Considering Fig. \ref{fig:tides_orbitele} in more detail, we see a clear wavy modulation in both the change of $a$ and $e$, that is due
to tidal couplings between the (BH1,BH2)-binary and the single BH3. The corresponding time-evolving modulation period directly
corresponds to the 1$\PN$+2$\PN$ orbital precession period of the (BH1,BH2)-binary.
The reason is that the tidal coupling between the (BH1,BH2)-binary and the BH3-single not only depends on the the BBH SMA relative to the
Hill sphere (bottom panel), but also on the BBH orbital orientation relative to the perturber, which to leading order will change on a time-scale
equivalent to the BBH precession period \citep[see also][]{2019MNRAS.487.5630H, 2019PhRvD.100d3010S}.
\\
\\
{\it The 1$\PN$+2$\PN$ orbital precession period is therefore imprinted in the time evolving orbital elements, that
propagates to the inspiral time of the binary through changes in the GW energy radiated at pericenter, which finally
shows up as an observable GW phase shift.}
\\
\\
Besides the observational opportunities of indirectly observing the wavy modulations in the orbital elements through the GW signal,
the most important indirect effect of the $\PN$-precession is essentially to average out any potential gradual energy exchange
from tides on the (BH1,BH2)-binary \citep[see e.g.][]{2019PhRvD.100d3010S}. In other words, this averaging effect from $\PN$-precession will protect the
binary from experiencing a run-away change in $a,e$ through, e.g., repeatedly expanding the binary through
tidal pumping over each orbit, which would otherwise result in a greater and greater unphysical large GW phase shift between the 3-body-perturbed and the
2-body-reference GW signal.
We highlight this, as the recent burst-timing-model put forward in \cite{2023PhRvD.107l2001R} does not include the 1$\PN$, 2$\PN$ precession terms,
which, as a result, leads to unphysically large GW phase shifts. In our Appendix we illustrate this directly, by showing simulations of a
system similar to the one presented in Fig. \ref{fig:Ex1_orbits}, but without the 1$\PN$, 2$\PN$ precession terms included. This results
in a clear systematic incorrect evolution of the binary and thereby GW phase shift.

Below we continue by considering how these tidal effects and the additional effect from Rømer delays give rise to modulations
in the GW peak frequency and a GW phase shift.

\subsection{GW Peak Frequency Modulations}\label{sec:GW Peak Frequency Modulations and Tides}

To explore the observable GW effects resulting from the presence of the perturbing BH3, we start by considering 
the top-panel in Fig. \ref{fig:Ex1_dt_dphi_etc}, which shows the GW peak frequency $f_p$ of the inspiraling (BH1,BH2)-binary approximated as 
\begin{equation}
f_p \approx \frac{1}{\pi}\sqrt{{Gm_{12}}/{r_p^{3}}},
\label{eq:fp_rp}
\end{equation}
where $m_{12} = m_1 + m_2$, and $r_p$ is the pericenter distance, as a function of time. The {\it blue} and the {\it orange} lines show results for the
3-body-perturbed and the 2-body-reference BBHs, respectively.
In the figure, each dot corresponds to the time of pericenter passage and illustrates at what points in time one would
see a GW burst when the source is still highly eccentric.
\\
\\
{\it One can interpret this time-series of pericenter-passages as a clock that represents the evolution of the binary
as a function of time. In the well-studied case of binary pulsars, one uses the time-modulation in the pulsar arrival times to infer
information about the system. Our `burst-clock' case is equivalent, as the GW burst modulations in both frequency and arrival time carry
imprints of the environment the binary inspirals within.}
\\
\\
The first feature we see in the GW peak frequency $f_p$ plot is (again) a clear wavy-modulation, that directly results
from the periodic tidal influence on the (BH1,BH2)-binary orbital elements from BH3, as we described in
Sec. \ref{sec:Tidal perturbations of Binary Elements}. Besides information from the GW phase shift, one can in principle also put further
observable constraints on the system if the GW peak-frequency $f_p$ modulation is observed; however, the modulation is naturally
largest at early times when the SMA of the binary is still large enough for the tidal couplings to be strong enough. It is therefore questionable
if these modulations have observational opportunities, as this will require accurately detecting the first few peaks of the signal. On the other hand,
one should note that the modulations in $f_p$ are independent of the observer location, whereas effects from, e.g., Rømer delay in principle
can be undetectable if the BBH is observed face-on relative to its orbit around the BH3-perturber.

\subsection{Rømer Delays Along the Orbit}\label{sec:GW Time Delays Along the Orbit}

We now consider the second panel in Fig. \ref{fig:Ex1_dt_dphi_etc}, which shows the difference in seconds, $\delta{t}$, in the time of arrival
of the GW signal emitted by the 3-body-perturbed BBH and the 2-body-reference BBH, respectively, as seen by the
{\it Observer} (see Sec. \ref{sec:Numerical Setup and Observers}). The time has here been shifted
such that $\delta{t} = 0$ at merger by definition (see also \citep{2017ApJ...834..200M}). Technically, for both the 3-body-perturbed binary and the 2-body-reference binary
we record the position and time of all pericenter passages (time and location of all peaks in $f_p$, which we consider as our `clock')
in the rest-frame of each respective binary, after which we propagate these time-series to the position of the
{\it Observer}. In the observer rest-frame we then take the time difference between
the time-series from the 3-body binary and the 2-body-reference binary, respectively, which gives us $\delta{t}$.
In other words, $\delta{t}$ is the light travel time (projected along the line of sight of the observer) between the 3-body-perturbed binary COM-trajectory
(curved trajectory) and the 2-body-reference binary COM-trajectory (linear trajectory) as a function of observer time, $t$, as also shown
in Fig. \ref{fig:BS_ill} (Note that $\tau$ in this figure denotes the actual light travel time and not the projected along the observer line of sight.
For our chosen position of the {\it Observer} these two timescales are near identical). This is also what is referred to as the Rømer delay.

The observed time difference $\delta{t}$ can in general be due to several physical effects. For illustrative purposes, we
split in the figure the computationally measured $\delta{t}$, denoted `Total' ({\it blue}), into its two contributions; tides ({\it grey}) and Rømer delay ({\it red}). The
`tides'-contribution arises from perturbations to the orbital elements of the 3-body binary that is
not present in the 2-body-reference binary (see Sec. \ref{sec:Tidal perturbations of Binary Elements}),
where the `Rømer delay'-contribution is due to the BBH inspirals along a curved orbit (around BH3) versus a straight line
(unperturbed). As seen in the second panel in Fig. \ref{fig:Ex1_dt_dphi_etc}, tides greatly dominate at early times of the inspiral,
which is expected as this is when the BBH is still wide enough to be in tidal contact with the third object.
The difference from the 2-body-reference evolution is large, which in principle would give rise to major hints
that something nearby is perturbing; however, observing the first few peaks
during the early stages when the eccentricity is still very high is extremely difficult, and will therefore not be the main interest of this paper.
Focusing at times closer to merger, we see that Rømer delay starts to dominate, essentially because the tidal contribution quickly fades away.
The actual time difference is getting smaller and smaller, but this is not necessarily the same as the GW phase shift, which is the quantity that
effectively tell us if these effects can be observed or not.
Finally, the oscillating behaviour in $\delta{t}$ is again a direct consequence of the 1$\PN$+2$\PN$-precession terms coupling to tides, hence
it is clear that $\delta{t}$ caused by tides does not just gradually build up as a function of time as otherwise argued in \cite{2023PhRvD.107l2001R} (see also our Appendix).

\subsection{GW Phase Shifts}\label{sec:GW Phase Shifts}

\begin{figure}
\centering
\includegraphics[width=0.5\textwidth]{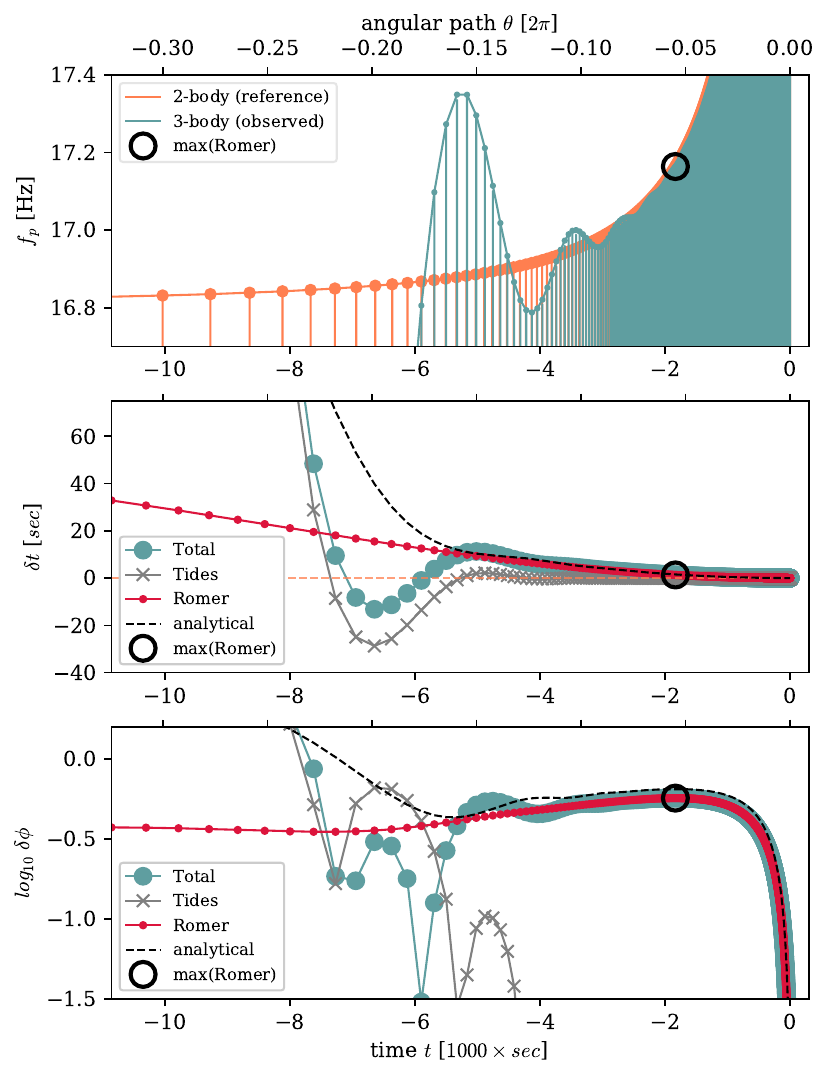}
\includegraphics[width=0.5\textwidth]{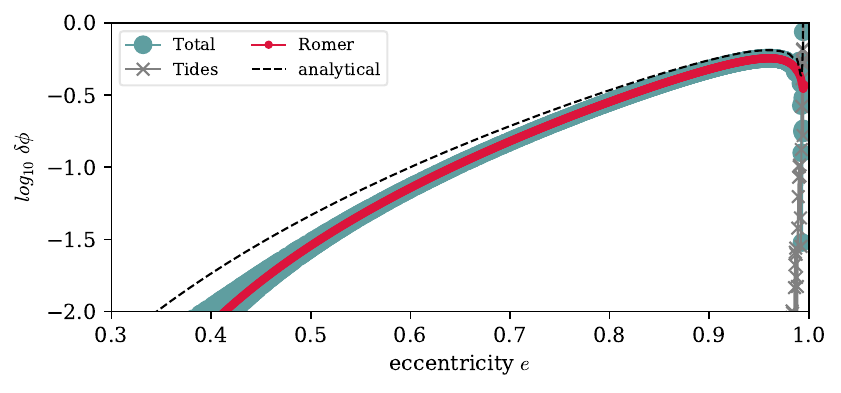}
\caption{{\bf Differences in observed GW signal:} Observable differences between the `3-body-perturbed' (BH1, BH2)-binary and
the `2-body-reference' binary shown in Fig. \ref{fig:Ex1_orbits}, and discussed in Sec. \ref{sec:Example and Numerical Methods}.
Counting from top to bottom:
{\it Panel 1.:} GW peak-frequency $f_p$ as a function of time for the real BBH (3-body (observed)) and the reference BBH
(2-body (reference)).
{\it Panel 2.:} Difference in arrival time seen by the {\it Observer}$, \delta t$, of the GWs sent out at each pericenter passage,
by the real orbiting binary and the reference binary, respectively, as a function of time $t$.
{\it Panel 3.:} The time-difference converted to GW phase shift, $\delta{\phi}$, as further described in Sec. \ref{sec:GW Phase Shifts}
and \ref{sec:General Relations}.
{\it Panel 4.:} GW phase shift, $\delta{\phi}$, as a function of the (BH1,BH2)-binary eccentricity.
The {\it red} and {\it grey} lines show the contribution from Rømer delay (time dependent Doppler shift) and dynamical tides (dynamical interactions),
respectively, where black dashed
lines show our full analytical solutions presented in Sec. \ref{sec:Analytical Models}.} 
\label{fig:Ex1_dt_dphi_etc}
\end{figure}

The third panel in Fig. \ref{fig:Ex1_dt_dphi_etc} shows the GW phase shift $\delta{\phi}$ between the 3-body binary and the 2-body-reference binary
as seen by the {\it Observer}. This can be approximated by the observed difference in number of orbital cycles
between the 3-body-perturbed signal and the 2-body unperturbed reference signal times $2\pi$, to convert the orbital count difference to
a shift in radians. As we illustrate later, this is indeed the phase shift of the GW-form in the limit where the (BH1,BH2)-binary is circular
(except for a trivial factor of two).
\\
\\
{\it In our numerical framework, we estimate the GW phase shift by taking the difference in arrival time
between the 3-body-perturbed- and 2-body-reference binary GW signals in the observer frame, $\delta{t}$, and divide this
by the corresponding orbital time of the 2-body-reference binary, $T(t)$, that we measure numerically from the $f_p$-peak
time locations. This fraction we then multiply by $2\pi$ to get the GW phase shift, such that $\delta{\phi} \approx 2\pi \delta{t}/T(t)$.}
\\
\\
This is also explained in Fig. \ref{fig:BS_ill}. Again, for illustrative purposes we have in Fig. \ref{fig:Ex1_dt_dphi_etc}
separated the contributions to $\delta{\phi}$ from tides and Rømer delay. Note, that the true GW phase
(connected to the evolution of the true anomaly that evolves non-linearly in time) is not easily measured when the binary is highly eccentric, in which case
the evolving GW signal is closer to a series of bursts. For this reason, we keep our definition of GW phase shift even when the binary is eccentric.
These technical issues will naturally not arise when working with real or NR generated wave-forms.

Considering further the results for $\delta{\phi}$ shown in Fig. \ref{fig:Ex1_dt_dphi_etc},
we see that if we track the binary backwards from merger ($t=0$) towards the time of assembly, the phase shift $\delta{\phi}$
first increases, as expected from earlier studies in the $e=0$ limit (\cite{2017ApJ...834..200M}), but does not continue to grow
as $e$ increases, i.e. as we go backwards in time.
Instead, it levels off and gradually decreases again, until it reaches a region where tides start to dominate
(the grey lines start fluctuating above the red line).
As we will analytically prove below, this shape of $\delta{\phi}$ is unique to eccentric GW sources orbiting around other objects.
In short, the reason why it levels off can be put as follows. First we note that
at a given time $t$, the binary will be at a given location on the outer orbit and thereby have a
given value for $\delta{t}(t)$, which is a purely geometric factor that has nothing to do with the evolution of the inner binary. However,
the orbital time of the inner binary, $T(t)$, that relates to $\delta{\phi}$ as $\sim \delta{t}/T(t)$, evolves by itself as a function of $t$ depending on
the properties of the binary and more generally on the dissipative force (gas-drag, GW radiation, etc.).
For GW radiation, the eccentric limit corresponds to a larger $T(t)$ at a given time $t$
compared to if the binary was circular, which then leads to a relatively smaller $\delta{\phi}$ compared to the circular case.
That the GW phase shift first increases (circular limit) after which it decreases (eccentric limit)
naturally gives rise to a peak in $\delta{\phi}$ in the Rømer delay dominated region, which we have highlighted with a black circle (Fig. \ref{fig:Ex1_dt_dphi_etc}).
This peak position is also shown in Fig. \ref{fig:Ex1_orbits} with a black circle, from which we see how close to merger the Rømer delay peak is
compared to the full orbital trajectory. The exact location of the peak naturally depends on the exact position of the observer;
What we show here is very close to the optimal case. The
trivial dependence on the observer position will not be addressed in this paper, as our main focus will be on first quantifying the intrinsic and unique
parts of the GW phase shift signal before the observer is factored in.

Finally, we have in the bottom panel of Fig. \ref{fig:Ex1_dt_dphi_etc} plotted the phase shift $\delta{\phi}$ as a function of the evolving BBH eccentricity, $e$.
(note that the binary here evolves from right to left). As seen, once the BBH eccentricity $e$ reaches a
value of $\approx 0.95$, the GW phase shift caused by Rømer delay is at its maximum. That the GW phase shift peaks at an eccentricity significantly
below the eccentricity at formation ($e_0 \sim 0.9999$), is encouraging, as the lower the eccentricity is, the less challenging the observations will be.
Below we explain all these results using analytical arguments, from which we will see how especially the GW phase shift
depends on time, eccentricity, mass- and length-scales.

\section{Analytical Solution}\label{sec:Analytical Models}

In this section we present the first analytical solution to the time-evolving GW phase shift resulting from Rømer delay, which relates to the non-relativistic Doppler shift,
of an eccentric BBH inspiraling along an orbit around a third BH.

\subsection{Assumptions and Setup}\label{sec:Assumptions and Setup}

We assume the outer orbit, i.e. the orbit of the BBH (GW source) around the single BH (perturber),
is circular to get a clear picture for how the mass-, length- and time-scales of the problem influence the magnitude and general evolution of $\delta{\phi}$.
The more general case with an eccentric outer orbit will be studied in upcoming work.
The setup and notations we use here are explained in Figure \ref{fig:BS_ill}. In this figure the center is at the COM of the 3-body system,
$R = R_{12} + R_3$ is the distance between the BBH and the single BH, which is a sum of the distance from the BBH, $R_{12}$, and the single BH, $R_3$,
to the 3-body COM, respectively. The inspiraling BBH has an initial SMA $a_0$, and eccentricity $e_0$, from which it evolves towards the point
of merger, which is denoted by `Merger' in the figure. Without loss of generality, we take the time at this point to be $t=0$, such that the time $t$
is equivalent to the merger time, $t_m$. To every value of $t$ there exists a value for the angle, $\theta$, which denotes
the angular evolution of the BBH around the 3-body COM w.r.t. the point of merger as illustrated in Fig. \ref{fig:BS_ill}. As was seen numerically in Fig. \ref{fig:Ex1_orbits},
the most important part of the GW phase shift caused by Rømer delay is created relatively close to merger, which is when $\theta$ is small,
and we will therefore in our analytical calculations assume that $\theta \ll 1$, as described further below.

Using this setup, we now calculate
the evolving {\it maximum} GW phase shift from Rømer delay, $\Delta{\phi}$, as described in Eq. \ref{eq:ADD_GWphs}.
With our choice of {\it Observer} (see Sec. \ref{sec:Numerical Setup and Observers}), the observed contribution
from Rømer delay in the limit $\theta \ll 1$ is identical to the maximum value, $\Delta{\phi}$, we derive here. The observed value $\delta{\phi}$
and the maximum value $\Delta{\phi}$ can therefore be compared directly as we will do in the following.

\subsection{General Relations}\label{sec:General Relations}

To estimate $\Delta{\phi}$ at each point along the trajectory of the BBH around the perturbing BH, we first imagine that we observe a GW burst
from the 3-body-perturbed BBH at some time $t$, with corresponding angle $\theta(t)$.
This GW burst is separated in time compared to a GW burst sent out from the
2-body-reference BBH by an amount $\tau$,
where $\tau$ is the light travel time between the 3-body-perturbed binary
COM position and the 2-body-reference binary COM position at time $t$ (pink lines in Fig. \ref{fig:BS_ill}). This
manifests as the Rømer GW phase shift. To see this, we continue by noting that
the distance between the 3-body and 2-body COM positions at time $t$, denoted here by $l$ as also shown
in Figure \ref{fig:BS_ill}, can be written as
\begin{align}
l 	& = 2{R_{12}} [\left({\theta}/{2}\right)^{2} + \sin^{2}\left({\theta}/{2}\right) \nonumber\\
	& - 2\left({\theta}/{2}\right)\sin\left({\theta}/{2}\right)\cos\left({\theta}/{2}\right)]^{1/2} \nonumber\\
	& \approx {2}R_{12}(\theta/2)^{2},\ (\theta \ll 1),
\end{align}
where in the last equality we have expanded to leading order in $\theta$ (our considered limit $\theta \ll 1$).
By now using that the angle $\theta$ can be written as,
\begin{equation}
\theta(t) = \frac{v_{12} t}{R_{12}},
\end{equation}
where $v_{12}$ is the COM velocity of the binary relative to the 3-body COM,
\begin{equation}
v_{12} = \sqrt{GM/{R_{12}}}
\end{equation}
we now find that the corresponding light travel time, $\tau = l/c$, can be written as,
\begin{equation}
\tau(t) = \frac{l(t)}{c} = \frac{1}{2} \frac{Gm_3}{c}\frac{t^2}{R^2}
\label{eq:tau}
\end{equation}
where we have used that $m_{12} v_{12} = m_3 v_3$, $m_{12} R_{12} = m_3 R_3$, and defined $M = m_1 + m_2 + m_3$.
This quadratic dependence on time $t$ is seen in Fig. \ref{fig:BS_ill}, where $\delta{t} \approx \tau$ given our choice of observer,
which fits our simulations when $\theta \ll 1$, or equivalently where $t$ approaches $0$ (black dashed line). Note that the
normalisation of $\tau$ scales as $1/R^2$, which indicates that the largest time-shifts at a given $t$ and in the $\theta \ll 1$ limit,
will be found for the most compact binaries.

With our derivation of $\tau$, we can now estimate the corresponding GW phase shift $\Delta{\phi}$, that can be written as
\begin{align}
	\Delta{\phi}(t) & \approx 2\pi {\tau(t)}/{T(t)}, \nonumber\\
	 	    & \approx \frac{1}{2}\frac{G^{3/2}}{c} \frac{m_3 m_{12}^{1/2}}{R^2} \times \frac{t^2}{a(t)^{3/2}}
    \label{eq:dphi_general}
\end{align}
where we have used Eq. \ref{eq:tau} and that the orbital time of the BBH at time $t$ can be estimated as,
\begin{equation}
T(t) = 2\pi\sqrt{a(t)^3/Gm_{12}}.
\end{equation}
As seen in Eq. \ref{eq:dphi_general}, the last term, $t^2/a(t)^{3/2}$, only depends on the binary properties,
and we can therefore conclude that $\Delta{\phi}$ will, regardless of the eccentricity of the binary, scale $\propto R^{-2}$.
As we are considering the $\theta \ll 1$ limit, this $R$ is essentially a measure of curvature, acceleration, or how much the orbital velocity vector
changes in time, i.e. $\dot{v}_{12}$ \citep{2011PhRvD..83d4030Y}.

If we assume that the BBH SMA, $a(t)$, scales with time as
$a(t) \propto t^{\alpha}$, one finds that
\begin{equation}
\Delta{\phi}(t) \propto t^{2-3\alpha/2},\ (a \propto t^{\alpha}),
\label{eq:dpi_general_atalpha}
\end{equation}
from which one can define a critical $\alpha$ defined as,
\begin{equation}
\alpha_c = 4/3, \ (a \propto t^{\alpha}).
\label{eq:alpha_crit}
\end{equation}
As seen, if $\alpha < \alpha_c = 4/3$ then $\Delta{\phi}(t)$ will increase with time $t$, whereas,
if $\alpha$ changes from being $< \alpha_c$ to $> \alpha_c$ along the BBH evolution, then $\Delta{\phi}$ will start to decrease.
This is exactly what we see in the case when the BBH transitions from its approximate circular state
($\alpha \approx 1/4$) around merger, to its earlier highly eccentric state ($\alpha \approx 2$), as further explained in the following.

\subsection{Evolution with Time}

To gain further analytical insight into $\Delta{\phi}$ as a function of time $t$, we have to consider the BBH $e=1$ and $e=0$ limits separately,
as no full analytical solution is formally possible as a function of $t(e)$.
A closed form analytical solution is on the other hand possible as a function of $e(t)$ as shown later in Sec. \ref{sec:Evolution with Eccentricity}.

\subsubsection{Circular Limit (e=0)}

Starting with the well studied $e=0$ limit (circular-orbit binary evolving around a perturber on a circular
orbit - \citep[see e.g.][]{2011PhRvD..83d4030Y, 2017ApJ...834..200M}), we first use the expression
for the merger time in the circular case $t_c$ from \cite{Peters64},
\begin{equation}
t_{c} = (5/256)(c^5/G^{3})a^{4}m_{1}^{-1}m_{2}^{-1}m_{12}^{-1},
\label{eq:tm_c}
\end{equation}
and substitute it into Eq. \ref{eq:dphi_general}, which leads us to,
\begin{equation}
\Delta{\phi}(t) \approx \frac{c^{7/8}G^{3/8}}{2} \left(\frac{5}{256}\right)^{3/8} \times \frac{m_3}{R^2}\frac{m_{12}^{1/8}}{m_1^{3/8}m_2^{3/8}} \times t^{13/8}.
\label{eq:dphi_e0_t}
\end{equation}
From this we see that $\Delta{\phi} \propto t^{13/8}$ ($e=0$), which explains the rapid increase in $\Delta{\phi}$ we see
in panel 3 in Fig. \ref{fig:Ex1_dt_dphi_etc}. Considering the case where $m_1 = m_2$, the above Eq. \ref{eq:dphi_e0_t} reduces to
$\Delta{\phi} \propto m_3R^{-2}m_{12}^{-5/8}t^{13/8}$. From this we see that $\Delta{\phi}$ generally is largest for high perturber mass $m_3$,
small outer orbital radius $R$, and small binary masses.

\subsubsection{Eccentric Limit (e=1)}

As we are going
backwards in time from the point of merger, the binary eccentricity increases until it reaches its initial eccentricity $e_0\sim1$.
We now consider the solution in this limit. For this we make use of the solution to the merger time in the eccentric limit from \cite{Peters64},
which can approximated by
\begin{equation}
t_e \approx t_c \times (768/425)(1-e^{2})^{7/2},
\label{eq:tm_ecc}
\end{equation}
where $t_c$ is given by Eq. \ref{eq:tm_c}.
By now using the identity $(1-e^2) = (1-e)(1+e)$, and the definition of the pericenter
distance, $r_{p} = a(1-e)$, the merger time can be written as
\begin{equation}
t_e \propto a^{1/2}r_{p}^{7/2},
\end{equation}
The radiation of angular momentum is
much smaller than the radiation of energy, which leads to the result that the pericenter distance remains nearly constant during the
eccentric phase of the inspiral. From this one concludes that the merger time in the $e \sim 1$ limit approximately
scales as $t_e \propto a^{1/2}$. This implies that in the $e=1$ limit, $\alpha$, defined in Eq. \ref{eq:dpi_general_atalpha}, takes the value
$\alpha = 2$ which leads to
\begin{equation}
\Delta{\phi}(t) \propto t^{-1}.
\label{eq:dphi_e1_t}
\end{equation}
This means that when the binary starts to have a significant eccentricity, the GW phase shift will start to decrease again after
having increased to a non-zero value during the $e\approx 0$ phase. This $\propto t^{-1}$ dependence is also confirmed using our simulations
shown in Fig. \ref{fig:Ex1_dt_dphi_etc}, as also discussed in Sec. \ref{sec:GW Phase Shifts}.
Note that we have not added any normalization and scaling factors in the above relation, as it has to be
bridged to the $e\approx0$ solution.
Below we illustrate how one can describe the GW phase shift $\Delta{\phi}$ and its scalings, across
the entire range of eccentricity $e$ by writing $\Delta{\phi}$ as a function of $e$ instead of $t$.

\subsection{Evolution with Eccentricity}\label{sec:Evolution with Eccentricity}

\begin{figure}
\centering
\includegraphics[width=\columnwidth]{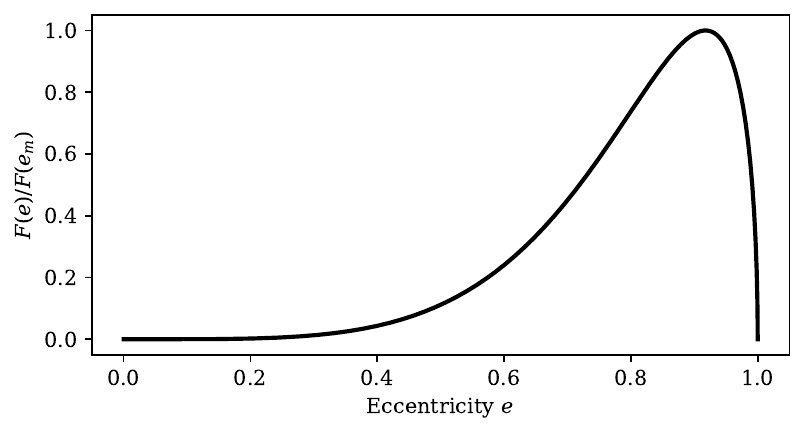}
\caption{{\bf GW phase shift and eccentricity:}
The figure pictures Eq. \ref{eq:FunctionFe}, that represents our analytically derived universal relation between the GW phase
shift, $\delta{\phi}$, and the eccentricity, $e$, of the BBH as it inspirals from its initial value $e_0 \sim 1$ all the way to merger at $e \sim 0$.
The curve has been normalized to its peak value, and illustrates therefore the general factor the GW phase shift scales by
when the eccentricity changes in time (here going from right to left). Interestingly, this scaling only depends on eccentricity, and not on the BH masses or
length scales. Note that the trivial factor from the position of the observer has not been factored in, the figure therefore illustrates an upper limit as a function
of eccentricity.
}
\label{fig:FeFem_e}
\end{figure}

To derive a solution for $\Delta{\phi}$ valid for all $e$ we start by recalling the relation between $a$ and $e$ given by \cite{Peters64},
\begin{equation}
a(e) = \frac{C_0e^{12/19}}{(1-e^2)} \times g(e),
\end{equation}
where
\begin{equation}
g(e) = \left(1+121e^2/304\right)^{870/2299},
\end{equation}
and $C_0$ is a constant that depends on
the initial conditions, $a_0, e_0$. In the case of dynamical assembly, the BBH will start with high eccentricity $e_0 \approx 1$,
from which it trivially follows that the constant $C_0 \approx 2r_{0}/g(1)$, where $r_0  = a_0(1-e_0)$. With this constant we can now write the relation $a(e)$ as, 
\begin{equation}
a(e) \approx \frac{2r_0e^{12/19}}{(1-e^2)}\frac{g(e)}{g(1)},\ (e_0 \approx 1).
\label{eq:ae_e1lim}
\end{equation}
By now substituting this $a(e)$ into our general expression for $\Delta{\phi}$ given by Eq. \ref{eq:dphi_general}, with $t$ being the merger time in the eccentric
limit from Eq. \ref{eq:tm_ecc}, one can write the following two useful expressions for $\Delta{\phi}$, as a function of $r_0$ and $f_0$, respectively, that depend only on $e$,
\begin{align}
	\Delta{\phi}(e) & \approx \frac{288\sqrt{2}}{85^{2}g(1)^{13/2}} \frac{c^{9}}{G^{9/2}} \times \frac{m_3}{R^2}\frac{r_0^{13/2}}{m_1^{2}m_2^{2}m_{12}^{3/2}} \nonumber\\
	 	    &  \times e^{78/19}(1-e^2)^{1/2}g(e)^{13/2} \nonumber\\
		    & \approx \frac{288\sqrt{2}\pi^{-13/3}}{85^{2}g(1)^{13/2}} \frac{c^{9}}{G^{7/3}} \times \frac{m_3}{R^2}\frac{m_{12}^{2/3}}{m_1^{2}m_2^{2}} \times f_{0}^{-13/3} \nonumber\\
	 	    &  \times e^{78/19}(1-e^2)^{1/2}g(e)^{13/2},
    \label{eq:dphi_e_e1lim}
\end{align}
where for the last relation we have used Eq. \ref{eq:fp_rp}.
As seen here, the dependence on $e$ factors out into the function,
\begin{equation}
F(e) = e^{78/19}(1-e^2)^{1/2}g(e)^{13/2},
\label{eq:FunctionFe}
\end{equation}
meaning that the functional shape of the GW phase shift $\Delta{\phi}$ as a function of $e$ has a unique form,
where the BH masses, outer orbital radius, and initial conditions for the BBH inspiral
only act as scaling factors to this general form. This solution is over-plotted with the numerical results shown
in Fig. \ref{fig:Ex1_dt_dphi_etc} in the bottom panel, from which we see excellent agreement. The function $F(e)$ itself is illustrated
in Fig. \ref{fig:FeFem_e}.
\\

{\it This analytical solution for $\Delta{\phi}(e)$ serves as one of our main results, as it
represents the first closed form solution to the question of what the GW phase shift is of an eccentric binary orbiting and merging around a third object.}

\subsection{Evolution with Frequency}\label{sec:Evolution with Frequency}

\begin{figure}
\centering
\includegraphics[width=0.5\textwidth]{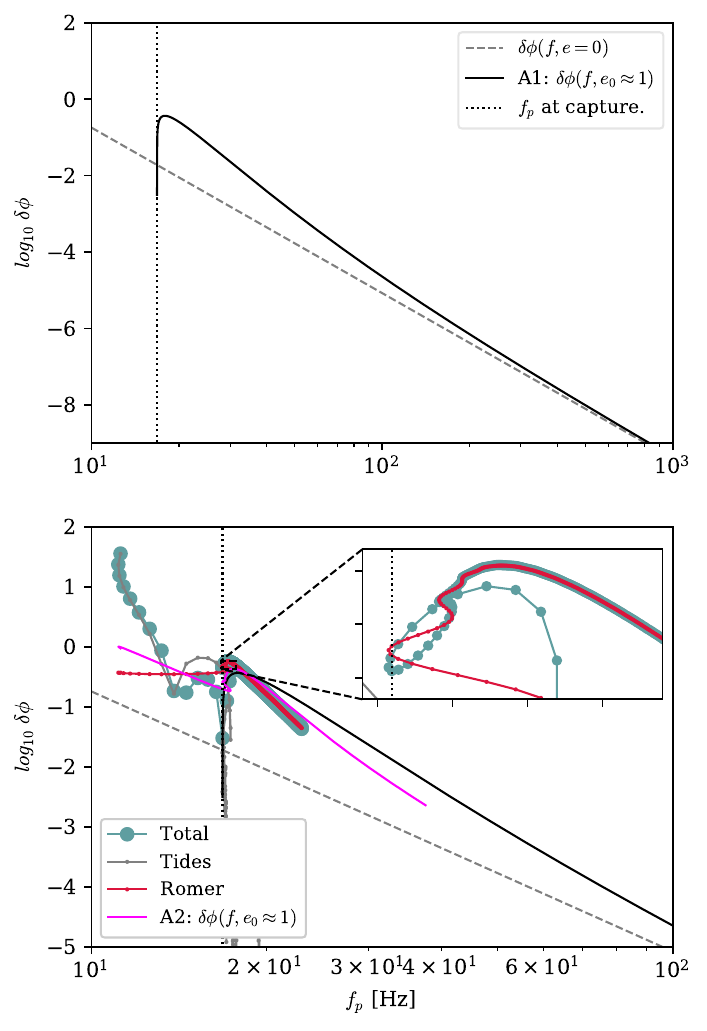}
\caption{{\bf GW phase shift and GW peak frequency:}
The figures show the GW phase shift, $\delta{\phi}$, as a function of GW peak frequency, $f_p$, for the $[5M_{\odot}, 5M_{\odot}] + 100M_{\odot}$
setup from Fig. \ref{fig:Ex1_orbits} and Fig. \ref{fig:Ex1_dt_dphi_etc}.
The lines corresponding to $\delta{\phi}(f, e=0)$ , A1: $\delta{\phi}(f, e_0 \approx 1)$ and A2: $\delta{\phi}(f, e_0 \approx 1)$ show the
analytical approximate solutions from Eq. \ref{eq:dphi_f_c}, Eq. \ref{eq:f_f_app} and Eq. \ref{eq:dphi_f_ecc_e}, respectively.
Note that Eq. \ref{eq:f_f_app} is the only one that can be written entirely as a function of $f$.
{\it Top:} Evolution of $\delta{\phi}$ from initial capture frequency $f_{0}$ (vertical dotted line) to where the eccentric approximate
solution ($\propto f^{-13/3}(1+({f_0}/{f}))^7(1-({f_0}/{f}))^{1/2}$) asymptotes to the classical circular solution ($\propto f^{-13/3}$).
As described in Sec. \ref{sec:Evolution with Frequency}, the eccentric evolution rises from its asymptotic scaling $f^{-13/3}$
to a peak that is generally $>10$ times larger than what is found in the circular case (see Eq. \ref{eq:H_max}).
{\it Bottom:} Zoom in on the evolution of $\delta{\phi}$ around the initial GW peak frequency, $f_0$, with the inclusion of the data from
the system shown in Fig. \ref{fig:Ex1_orbits}. The slightly more accurate solution from Eq. \ref{eq:dphi_f_ecc_e} is also included and
plotted using the geometric estimated eccentricity from the data (see Sec. \ref{sec:Tidal perturbations of Binary Elements}).
In general, it is clear that $\delta{\phi}$ quickly decreases with increasing $f$, which adds to our conclusion from Sec. \ref{sec:Peak Frequency at Maximum}
that the largest GW phase shift will happen at GW frequencies around the initial frequency, $f_0$.}
\label{fig:dphi_freq}
\end{figure}

It is useful to consider $\Delta{\phi}$ as a function of GW frequency, $f$,
as this is naturally closely connected to observational constraints. For example, if the GW source initially forms
outside the observable band, i.e. at lower frequencies,
one would like to know what the GW phase shift is once the source enters the band, i.e. passes some characteristic frequency.
To link our work with earlier studies, we first write out $\Delta{\phi}$ as a function of $f$ assuming the BBH is inspiraling with $e=0$ during
its entire evolution. Using Eq. \ref{eq:fp_rp} and Eq. \ref{eq:tm_c}, we can now rewrite Eq. \ref{eq:dphi_e0_t} as,
\begin{equation}
\Delta{\phi}(f,e=0) \approx \frac{c^{9}G^{-7/3}}{2\pi^{13/3}} \left(\frac{5}{256}\right)^{2} \times \frac{m_3}{R^2}\frac{m_{12}^{2/3}}{m_1^{2}m_2^{2}} \times f^{-13/3}.
\label{eq:dphi_f_c}
\end{equation}
This is exactly the same expression as the one derived in \cite{2017ApJ...834..200M} (Sec. 4.1),
including the prefactor, that, in the same units and notation as in \cite{2017ApJ...834..200M},
here takes the form $8\pi^{-13/3}(5/256)^{2}c^{9}G^{-7/3}M_{\odot}^{-7/3}R_{\odot}^{-2}10^{-13/3} \approx 446$ (except for an expected
factor of $2$, as we consider the phase shift in terms of binary orbital periods to make
it possible to smoothly transition from the eccentric to the circular limit, where \cite{2017ApJ...834..200M} naturally define
it in terms of GW cycles). This consistency serves as an excellent check of our entire approach and numerical framework.

In the general $e>0$ case, a closed form solution between $\Delta{\phi}$ and $f$ is impossible, as $e$ cannot be written as a function of $f$,
as we will explain below. However, we can still get insight into this after a few approximations. For this, we start by combining
Eq. \ref{eq:ae_e1lim} and Eq. \ref{eq:fp_rp}, to find how the GW peak frequency changes as a function of eccentricity as the binary spirals in,
\begin{equation}
\left({f}/{f_0}\right)^{2/3} \approx \frac{(1+e)}{2e^{12/19}}\frac{g(1)}{g(e)}, 
\label{eq:f_ecc}
\end{equation}
where $f_0$ denotes the GW peak frequency at assembly. By now taking this expression to the power of $13/2$, one can rearrange the above
equations as $f^{-13/3}(1+e)^{7}2^{-13/2} = {f_0}^{-13/3}e^{78/19}(1+e)^{1/2}\left(g(e)/g(1)\right)^{13/2}$. This relation can now be substituted
into Eq. \ref{eq:dphi_e_e1lim}, from which we find,

\begin{equation}
\Delta{\phi}(f) \approx \Delta{\phi}(f,e=0) \times H(e),
\label{eq:dphi_f_ecc_e}
\end{equation}
where $\Delta{\phi}(f,e=0)$ is given by Eq. \ref{eq:dphi_f_c}, and $H(e)$ is given by,
\begin{equation}
H(e) = (1+e)^7(1-e)^{1/2}.
\label{eq:He_fac}
\end{equation}
Ultimately, we wanted this to be as a function of frequency $f$ only, but $e$ cannot be
isolated algebraically as a function of $f$. However, the above relations tell us that in general the GW phase shift of an eccentric binary does
not take a simple power law form as a function of $f$, instead, it will have unique features that depends strongly on the binary eccentricity,
$e$, which is connected to the frequency evolution. More specifically, when going backwards from
the point of merger the term $H(e) = (1+e)^7(1-e)^{1/2}$ will increase until $e$ reaches the value $e_H = 13/15 \approx 0.9$ (the value that maximizes $H(e)$),
after which $H(e)$ will strongly fall towards $0$ when $e = e_0 \sim 1$, i.e. when $f \sim f_0$. That is, the GW phase shift will increase faster than the
circular $f^{-13/3}$ when the binary is eccentric, until $e\sim 0.9$, after which it quickly declines
(note that where $H(e)$ peaks is not necessarily where $\Delta{\phi}$ peaks as will be clear in Sec. \ref{sec:Maximum GW Phase Shift} below).
At its maximum
\begin{equation}
H(e_H) = (1+13/15)^7(1-13/15)^{1/2} \approx 30, 
\label{eq:H_max}
\end{equation}
which tells us that for most of the inspiral, the GW phase shift of an eccentric binary
will be {\it larger} than a circular binary with the same GW frequency, by up to a factor of $\sim 30$ at its peak. If this means
that future 3G GW detectors are more sensitive to GW phase shifts in such inspirals will be topics of future
work, as the correct SNR has to be factored in.

A crude approximation makes it possible to get a rough estimate for how $\Delta{\phi}(f)$ scales with frequency only, without the eccentricity dependence
as we have above. To see this, we consider the $e \approx 1$ limit for which we can write
Eq. \ref{eq:f_ecc} as $(f/f_0)^{-2/3} \approx e^{12/19}$, that leads to
$e \approx (f_0/f)^{19/18} \approx (f_0/f)$. Now plugging this into Eq. \ref{eq:dphi_f_ecc_e},
we find the following approximate relation,
\begin{equation}
\Delta{\phi}(f) \approx \Delta{\phi}(f,e=0) \times (1+({f_0}/{f}))^7(1-({f_0}/{f}))^{1/2}.
\label{eq:f_f_app}
\end{equation}
Fig. \ref{fig:dphi_freq} shows our considered different solutions, together with the results from the system introduced in Fig. \ref{fig:Ex1_orbits}.
As seen, our approximate analytical solutions (Eq. \ref{eq:dphi_f_ecc_e} (pink) and Eq. \ref{eq:f_f_app} (black)) correctly captures
both the shape and the normalisation compared to the numerically estimated results. This encourages using
the simple relation Eq. \ref{eq:f_f_app} for doing quick estimates for how large the expected GW phase shift can be for different
dynamically assembled systems as a function of GW frequency, $f$. On the technical side, one sees
in Fig. \ref{fig:dphi_freq} that the GW phase shift from the data has been cut-off in this example at about $25$Hz.
The reason is simply that $\Delta{\phi}$ drops so fast with increasing $f$, that it poses numerical challenges in resolving how the
system asymptotes to the circular limit (grey dashed line) when starting from highly eccentric initial conditions. We did several checks with very
fine-tuned systems to ensure that indeed we approach the circular limit $\propto f^{-13/3}$.
From this we conclude that the largest chance of seeing the GW phase shift is clearly
when then system is close to its initial GW frequency, as $\Delta{\phi}$ quickly fades away with increasing $f$. This is further discussed below.

\subsection{Maximum GW Phase Shift}\label{sec:Maximum GW Phase Shift}

The eccentricity for which the GW phase shift will have its peak value, i.e. where $\Delta{\phi}(e_m) = \max(\Delta{\phi}(e))$,
can be estimated by deriving the eccentricity that maximizes the function $F(e)$ from Eq. \ref{eq:FunctionFe}. From this we find
\begin{equation}
e_m = \sqrt{\frac{2\left(\sqrt{391681} - 115 \right)}{1213}} \approx 0.92
\label{eq:e_maxphi}
\end{equation}
Note that this value is slightly lower than the one we find using our full $\PN$ $N$-body simulations shown in Fig. \ref{fig:Ex1_dt_dphi_etc}, which is closer to
$\sim 0.95$. This is partly due to our approximations such as the merger time in the eccentric limit that will, as the eccentricity decreases,
end up by a factor of $768/425$ larger that the true $t_c$ when $e$ approaches $0$. These differences are not of any concern here (and could be avoided
if one uses fitting functions such as the one presented in \cite{2020MNRAS.495.2321Z}); our main result
is that $\Delta{\phi}(e)$ has a unique form given by Eq. \ref{eq:FunctionFe} with a peak ($F(e_m) \approx 0.5$) that surprisingly is
located at a universal value for the eccentricity ($e_m \sim 0.95$) regardless of the BH masses and orbital length scales.
Therefore, although the source spends a significant amount of time in its eccentric state right after assembly, the maximum $\Delta{\phi}$
is not reached until $e\sim 0.95$.
The GW phase shift, relative to the maximum value, does fall off relatively quickly when the eccentricity decreases towards merger. This is
naturally given by Eq. \ref{eq:FunctionFe}, which scaled to its maximum value is,
\begin{equation}
\frac{\Delta{\phi}(e)}{\Delta{\phi}(e_m)} = \frac{F(e)}{F(e_m)} \approx 2e^{78/19}(1-e^2)^{1/2}g(e)^{13/2}.
\label{eq:phiphimax}
\end{equation}
For example, when the source has decreased to $e\approx0.5$, ${\Delta{\phi}(e=0.5)}/{\Delta{\phi}(e_m)} \approx 10^{-1}$ and $\approx 10^{-2}$ at
$e\approx 0.3$, which implies that for both current and future detectors, the GW phase shift is only expected to have a significant value when the
inspiraling BBH still is in its eccentric stage with $e \gtrsim 0.5$. The above relation is further shown in Fig. \ref{fig:FeFem_e}. 
Further discussions of SNR-calculations of eccentric binaries including Rømer delay and tidal effects are reserved for a future study,
but a `rule-of-thumb' is that a LIGO-like detector can resolve GW phase shifts of order $1$ radian
$\times$  $8/\text{SNR}$ \cite[e.g.][]{2017ApJ...834..200M}, which offers an extremely
promising future for GW phase shift detections from eccentric BBHs observed by 3G GW detectors, where SNR of order $\sim 1000$ is expected for a significant
fraction of the sources.

\subsection{Peak Frequency at Maximum}\label{sec:Peak Frequency at Maximum}

A relevant question relates to what the GW peak frequency is when the BBH reaches its maximum GW
phase shift, denoted here by $f_m$, and how this is related to the initial GW peak frequency, $f_0$.
To see this we first use Eq. \ref{eq:ae_e1lim} and rewrite it in terms of pericenter distance,
\begin{equation}
r_p(e)/r_0 \approx {2e^{12/19}}{(1+e)^{-1}}{g(e)}{g(1)^{-1}}.
\end{equation}
This now leads to the following expression for the ratio between the initial GW peak frequency, $f_0$ ($e_0=1$), and the GW peak frequency leading to
maximum $\Delta{\phi}$, $f_m$ ($e=e_m$),
\begin{equation}
\frac{f_{m}}{f_0} \approx \left( \frac{2e_m^{12/19}}{1+e}\frac{g(e_m)}{g(1)} \right)^{-3/2} \approx 1\ (e_m \approx 0.95).
\label{eq:fmf0_e}
\end{equation}
From this we conclude that $f_m \sim f_0$, which means that the GW peak frequency the inspiraling BBH will
have, when $\Delta{\phi}$ is at its maximum, will be very close to the GW peak frequency it had at formation, $f_0$.
This leads to the conclusions we also had above, namely that if a BBH has to be observable near its maximum GW phase shift,
the BBH has to form with an initial GW peak frequency in the observable band, or at least very close to.

\subsection{Dynamical Constraints}\label{sec:Dynamical Constraints}

One question remains, namely, how large can the GW phase shift $\Delta{\phi}$ theoretically get if we impose dynamical constraints on the input parameters?
From Eq. \ref{eq:dphi_e_e1lim}, we see a relatively strong dependence on the initial pericenter distance $r_0$, which could indicate that the GW phase shift can get extremely large,
if we just allow for a large enough value of $r_0$. However, $r_0$ cannot take any value. The reason is simply that $r_0$, and the corresponding GW
radiation emitted over one passage at this distance, is linked to the corresponding value of $a_0$, which has to be within the BBH Hill sphere, imposed by the
perturber, $R_H$.
To see these connections, we first note that as the BBH consisting of $m_1, m_2$ essentially `forms' through a GW radiation assembly, then after
the first pericenter passage the BBH orbital energy,
\begin{equation}
E_{0} = \frac{Gm_1m_2}{2a_0},
\end{equation}
will approximately equal the energy radiated
over the first passage,
\begin{equation}
\Delta{E}_{0} \approx \frac{85{\pi}G^{7/2}}{12\sqrt{2}c^{5}} \frac{m_1^2m_2^2\sqrt{m_{12}}}{r_0^{7/2}}.
\label{eq:DE_GW}
\end{equation}
By now equating $E_{0}$ and $\Delta{E}_{0}$, we can express the initial SMA $a_0$ that follows from the first passage at pericenter $r_0$ as,
\begin{equation}
a_0 \approx \frac{6\sqrt{2}}{85\pi}\frac{c^{5}}{G^{5/2}}\frac{r_0^{7/2}}{m_1m_2m_{12}^{1/2}}.
\label{eq:a0_r0}
\end{equation}
For the BBH to evolve after this first passage without being torn apart by the tidal forces of the third object with mass $m_3$ at distance $R$, its
SMA $a_0$ has to be less than the Hill sphere given by Eq. \ref{eq:RH}.
In this setup the maximum value for
$a_0$ will therefore be $\sim R_H$, which now allow us to solve for the corresponding maximum $r_0$ using
Eq. \ref{eq:a0_r0}, from which we find
\begin{equation}
r_H \approx \left(\frac{85\pi}{6\sqrt{2}} \frac{G^{5/2}}{c^{5}} \frac{Rm_1m_2m_{12}^{5/6}}{m_3^{1/3}}\right)^{2/7}.
\label{eq:rH_RH}
\end{equation}
Here the subscript $H$ refers to that this initial pericenter distance will result in a BBH with initial SMA $a_0$
that is about the size of the Hill sphere, $R_{H}$. Note that a few standard approximations are introduced in the estimation for the Hill sphere,
and the value for $r_H$ is therefore only approximate; however, it should carry the relevant scalings of the problem.
The maximum GW phase shift in our setup can now be found be substituting $r_0$ in Eq. \ref{eq:dphi_e_e1lim} with $r_{H}$ from Eq. \ref{eq:rH_RH},
after which we find,
\begin{align}
	\Delta{\phi}_H(e) & \approx \frac{288}{g(1)^{13/2}}\left(\frac{\pi^{13}}{6^{13}680}\frac{G}{c^2}\right)^{1/7} \nonumber\\
			 & \times	\left(\frac{m_3}{R}\right)^{1/7} \left(\frac{m_3m_3}{m_1m_2}\right)^{1/7} \left(\frac{m_{12}}{m_3}\right)^{1/21} F(e) \nonumber\\
			 & \approx 1 \times \left(\frac{m_3/M_{\odot}}{R/AU}\right)^{1/7} \left(\frac{m_3m_3}{m_1m_2}\right)^{1/7} \left(\frac{m_{12}}{m_3}\right)^{1/21} F(e).
    \label{eq:maxphi_rH}
\end{align}
We first note that the normalization is encouraging, as it suggests that the characteristic value for the maximum GW phase shift is about $1$ radian.
Furthermore, it is interesting that it does not depend strongly on neither the BH masses $m$ nor the orbital distance $R$, due to
the corresponding extremely low power-factors of $1/21$ and $1/7$, respectively. A maximum of $\sim 1$ radian for eccentric sources therefore seems
like a universal upper limit according to our theoretical models, and for the setups we here consider.

However, as discussed in Sec. \ref{sec:Numerical Setup and Observers},
the tidal influence from the BH3 perturber during the first few cycles of the BBH right after GW assembly,
can greatly affect the initial orbital conditions, which easily can bring the BBH into a regime that leads to an orbital evolution that
exceeds the upper-limit of $\Delta{\phi}$ derived above. Another point is that our considered Hill sphere estimate only
applies to interactions that are near co-planar with the orbit around the perturber, as the tidal forces that
give rise to what is defined as the Hill sphere change with the orbital orientation of the binary relative to the orbital plane
around the perturber. This is clear from considering the tidal tensor, which, e.g., gives rise to the well-known tidal
compression in tidal disruption events that are opposite to the tidal expansion in the plane \citep[e.g.][]{Carter:1982fn}. 
To conclude, the tidal influence from the perturber can lead to a system with an effective pericenter distance
that is $>r_H$ and thereby a $\Delta{\phi}$ that exceeds our derived limit, which indicates
that observing GW phase shifts are possible under the right (observer) conditions.

To further illustrate the promising future of GW phase shift sources, we conclude this study by
considering two astrophysical examples. The first being the GW capture and merger of two BHs in an AGN migration
trap around a SMBH, where the second example shows results from a chaotic 3-body interaction taking place inside a GC.

\section{Astrophysical Examples}

\begin{figure}
\centering
\includegraphics[width=\columnwidth]{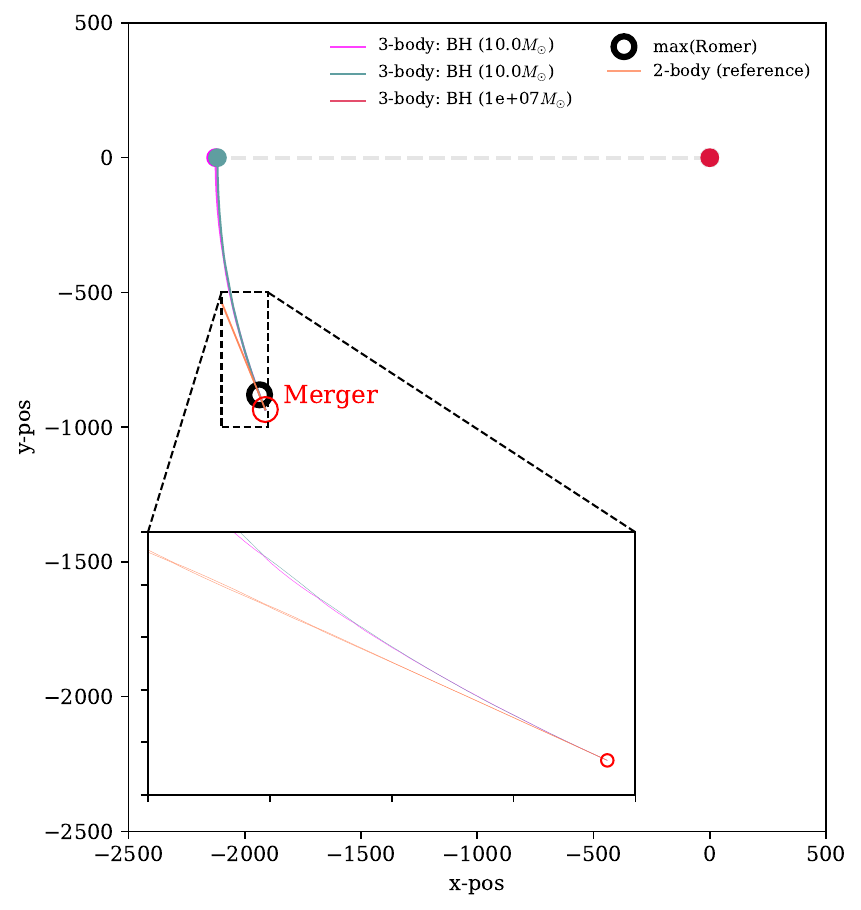}
\includegraphics[width=\columnwidth]{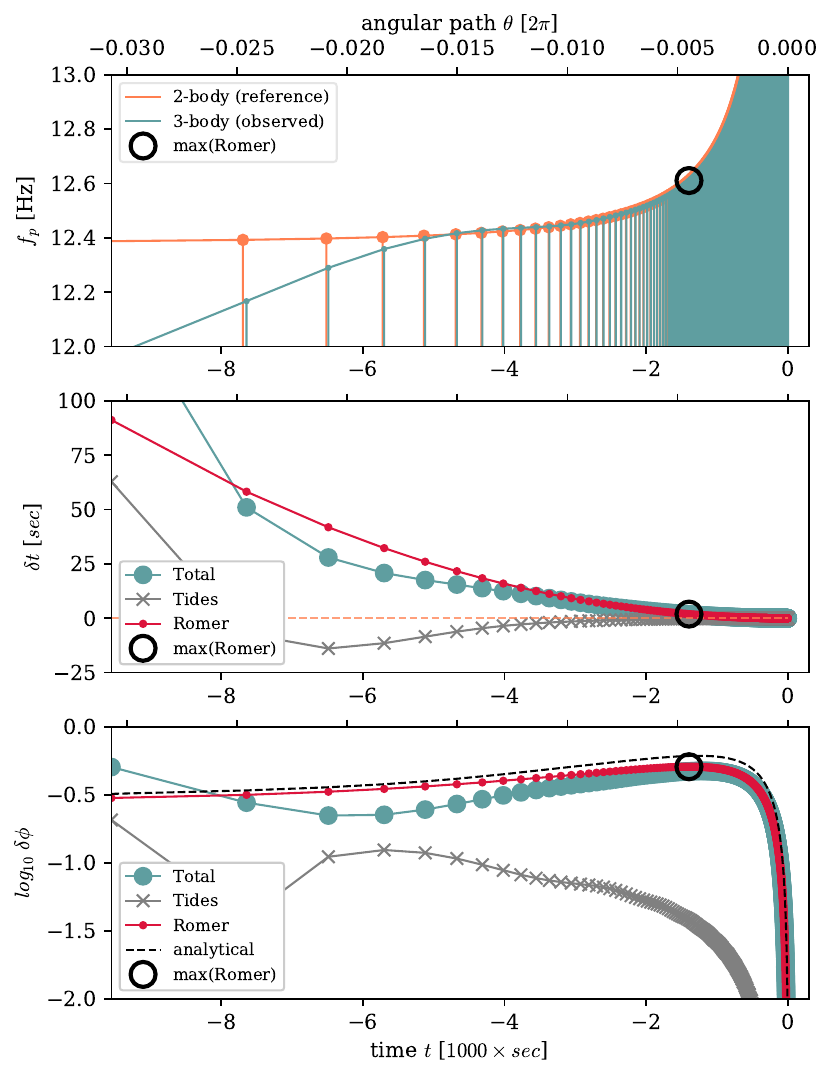}
\caption{{\bf GW merger in an AGN migration trap}. Observable differences between the 3-body-perturbed binary ($m_1 = m_2 = 10M_{\odot}$) orbiting
a central SMBH ($m_3 = 10^{7}M_{\odot}$) at a distance of $50$ Schwarzschild radii, and the 2-body-reference binary,
for the setup shown in the top panel and further described in Sec. \ref{sec:GW merger in an AGN migration-trap}.} 
\label{fig:AGNtrapEx}
\end{figure}

The formation and properties of GW inspirals with promising GW phase shifts can be rather chaotic, at least when it comes to the
normalization of the GW phase shift, as this strongly depends on the initial conditions and the following tidal couplings to the inspiraling BBH during the first few orbits.
We will quantify the distribution of GW phase shifts from different BBH merger channels using statistical methods in upcoming papers,
but here we conclude this paper by presenting two astrophysically relevant cases that are known, at least theoretically, to be able to create
eccentric BBH mergers with a perturber nearby.

\subsection{GW Merger in an AGN Migration-Trap}\label{sec:GW merger in an AGN migration-trap}

BBHs captured or formed in AGN disks will migrate through the disk towards the central SMBH \citep[e.g.][]{2020ApJ...898...25T}. Along their way they might undergo
pairings and subsequent few-body interactions that very likely lead to mergers \citep[e.g.][]{2022Natur.603..237S}. Depending on the AGN disk profile and
properties, there could be so-called AGN trap regions where the migration forces switch sign giving rise to regions
where BHs pile up \citep[e.g.][]{2016ApJ...819L..17B}. These are known as `migration traps' and could naturally facilitate a pit for bringing BHs to merger.
As described in e.g. \cite{2016ApJ...819L..17B}, migration traps can be found at distances as close as
$\sim 50 \times \mathscr{R}_3$ from the SMBH, where $\mathscr{R}_3$ denotes the Schwarzschild radius of the SMBH (BH3).
The location and even existence of these AGN migration traps are still under debate from
a theoretical perspective, but GW observations can here help constraining these models, as we show here.

We consider a BBH ($m_1= m_2 = 10M_{\odot}$) undergoing a GW assembly and subsequent inspiral around a SMBH ($m_3 = 10^{7}M_{\odot}$),
at a distance $R = 50 \mathscr{R}_3$. The evolution and resulting GW phase shift for this system is shown in Fig. \ref{fig:AGNtrapEx}.
As seen, the BBH has an initial GW peak frequency near the LVK-band and a
corresponding maximum GW phase shift approaching $1$ radian. Such BBH mergers forming in AGN traps
therefore have the potential to be observed even with current detectors, and certainly with future 3G detectors,
which encourages more detailed work on such GW phase shift effects in BBHs forming dynamically
in AGN disk environments.

\subsection{Chaotic 3-body Scattering Merger}\label{sec:Chaotic 3-body Scattering Merger}

\begin{figure*}
\centering
\includegraphics[width=\textwidth]{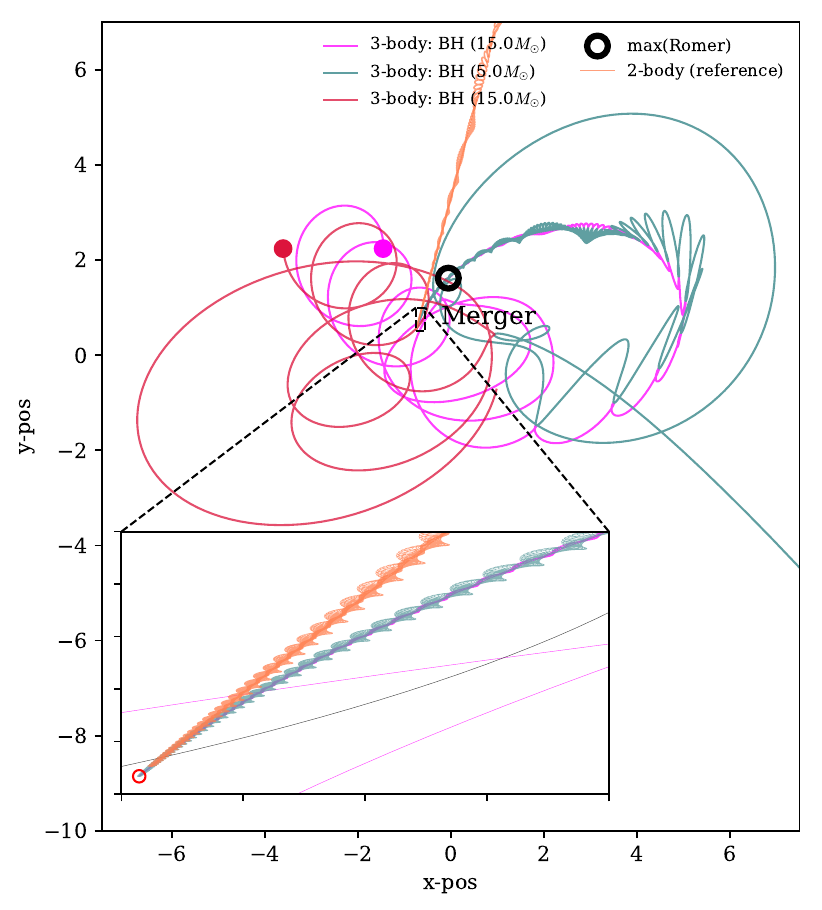}
\caption{{\bf Chaotic 3-body Scattering Merger.}
Formation of an eccentric GW source during an interaction between a ($15M_{\odot}$, $15M_{\odot}$) BBH (red and pink)
interacting with an incoming $5M_{\odot}$ BH (blue).
As seen, the final merger is between one of the $15M_{\odot}$ BHs (pink) and the incoming $5M_{\odot}$ BH (blue). During the GW inspiral, the GW-form seen
by a distant observer will be perturbed by the presence of the third bound object (red). This perturbation can be used to not
only probe the existence of the third object, and thereby the underlying formation mechanism, but also the masses and orbits of the full three-body
system as in the case of binary pulsars. We quantify the perturbations by comparing the `2-body-reference' binary (orange) with the real `3-body-perturbed' binary, as
further explained in Sec. \ref{sec:Gravitational Wave Phase Shifts} and \ref{sec:Chaotic 3-body Scattering Merger}.
Corresponding time- and GW phase shift results are shown in Fig. \ref{fig:GW3bodyorbit_PHASEetc}. 3-dimensional
Stereographic Images of this scattering is shown in Fig. \ref{fig:3dviews}.}
\label{fig:GW3bodyorbit_ill}
\end{figure*}

\begin{figure*}
\centering
\includegraphics[width=\textwidth]{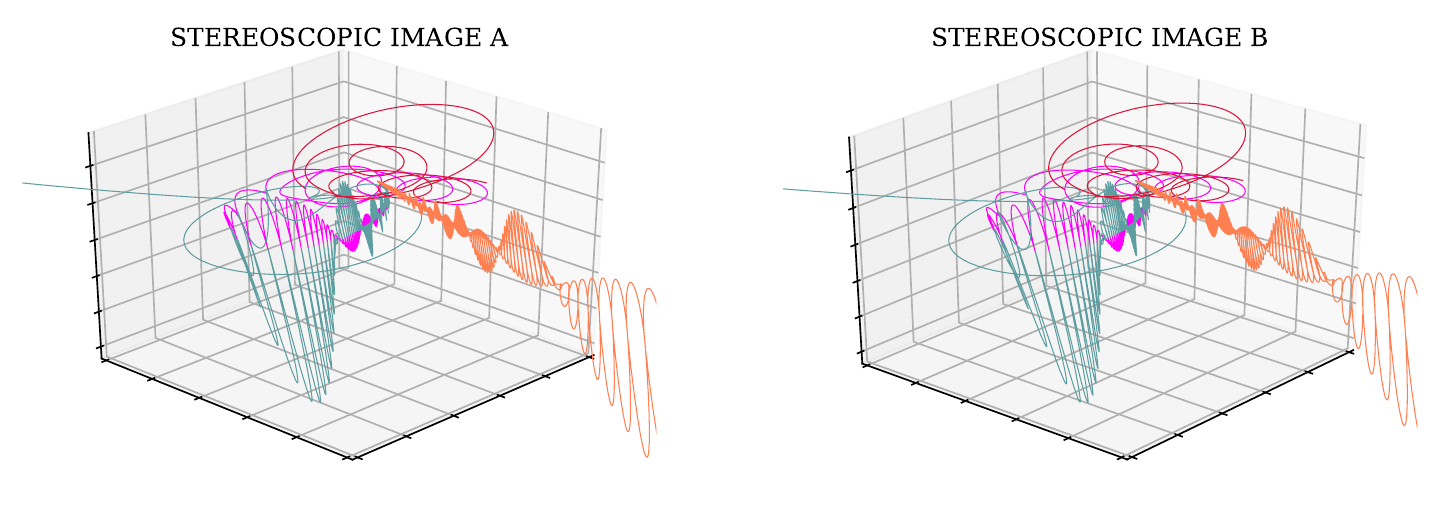}
\includegraphics[width=\textwidth]{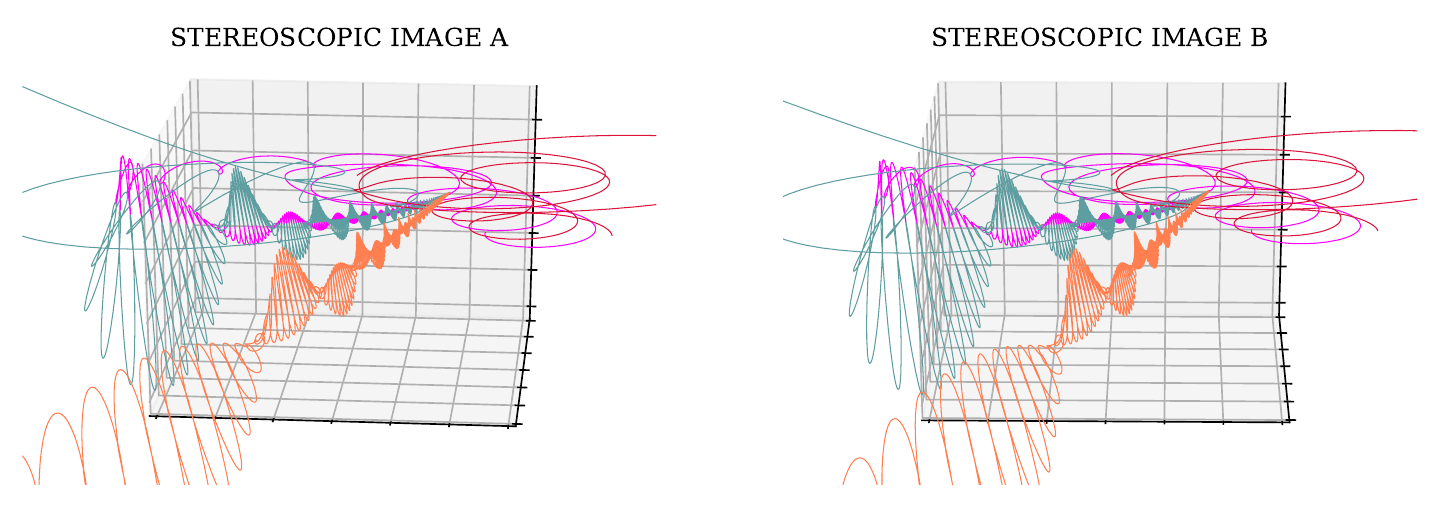}
\includegraphics[width=\textwidth]{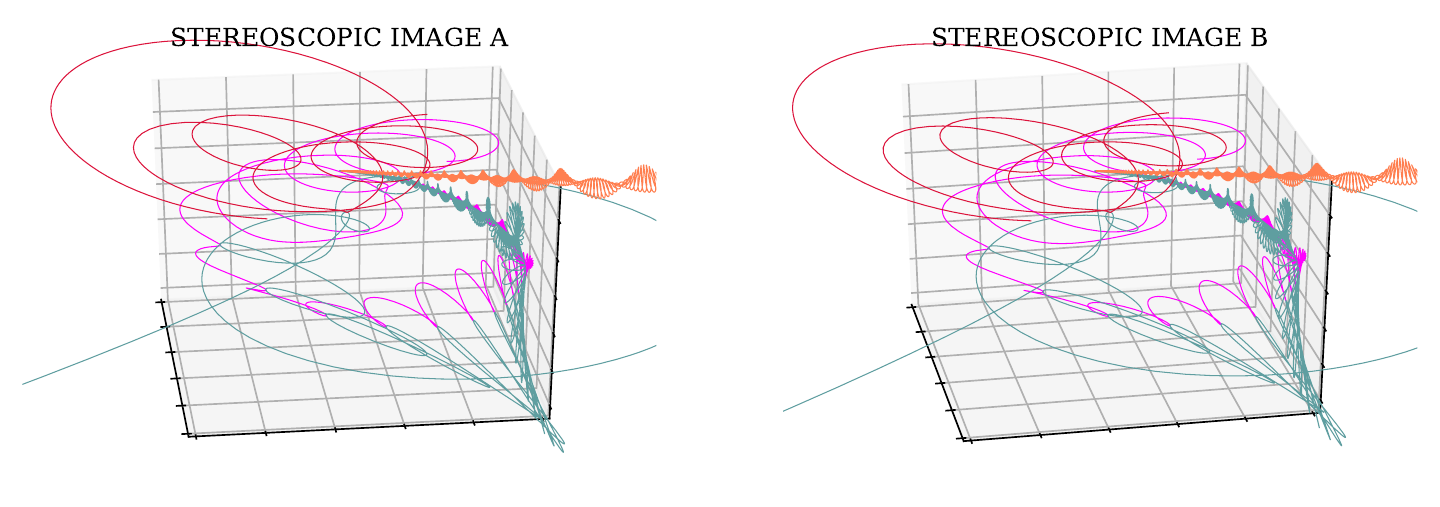}
\caption{{\bf Stereoscopic 3-Dimensional Images.}
The three horizontal panels each shows 3D plots of the chaotic 3-body scattering from Fig. \ref{fig:GW3bodyorbit_ill}, to provide a 3-dimensional
understanding for how and where the 2-body-reference binary is in relation to the 3-body-perturbed binary. To see and experience the
3-dimensional effect, one has to visually overlap `STEREOSCOPIC IMAGE A' with `STEREOSCOPIC IMAGE B' using a cross-eyed view.
{\bf This is how you do it}: Look normally at a point right between image `A' and `B'. Now take your finger up in front of your eyes along your line of sight.
While keeping your focus on the point right between image `A' and `B', move your finger back and forth until the two images you see of your finger visually
are `on top' and in `the middle' of each image `A' and `B'. Now shift your focus to your finger. Without shifting focus, remove your finger, and make a slight
adjustment to now see image `A' and `B' overlap to form one 3-dimensional picture popping out of the screen/paper.
With a bit of practice you will be able to quickly see these
stereoscopic images while your fingers are safely placed in your pocket.}
\label{fig:3dviews}
\end{figure*}

\begin{figure}
\centering
\includegraphics[width=\columnwidth]{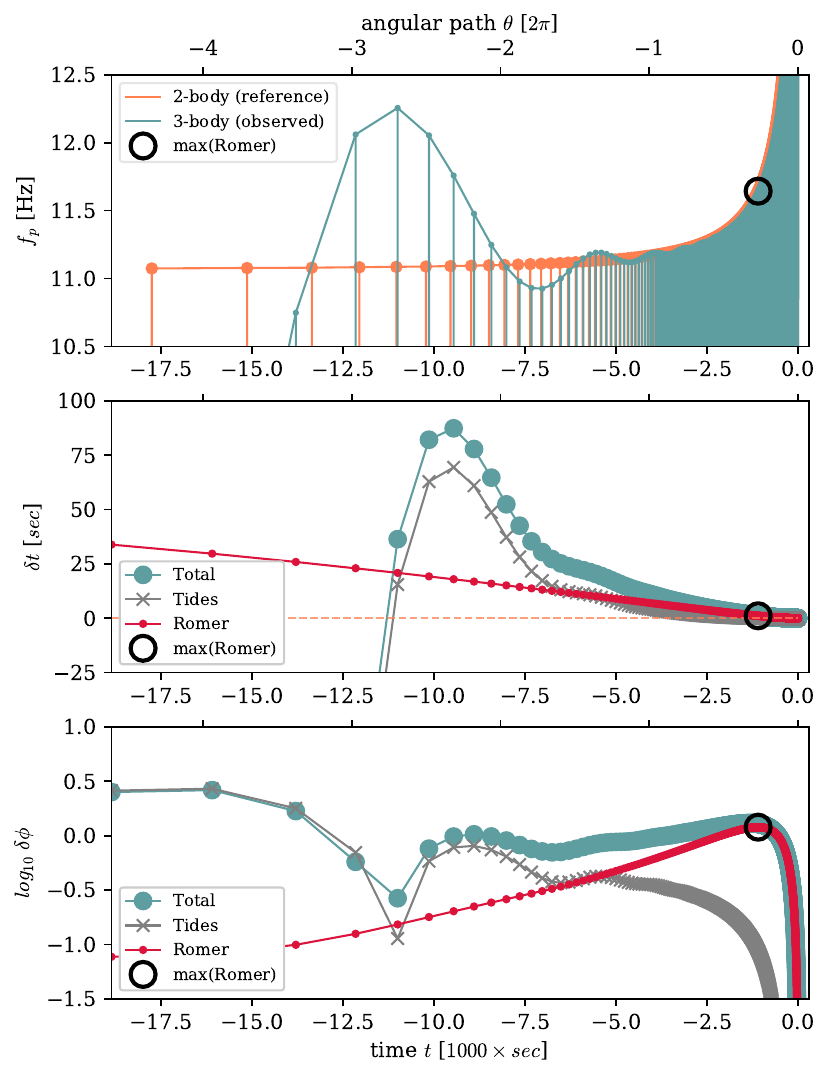}
\includegraphics[width=\columnwidth]{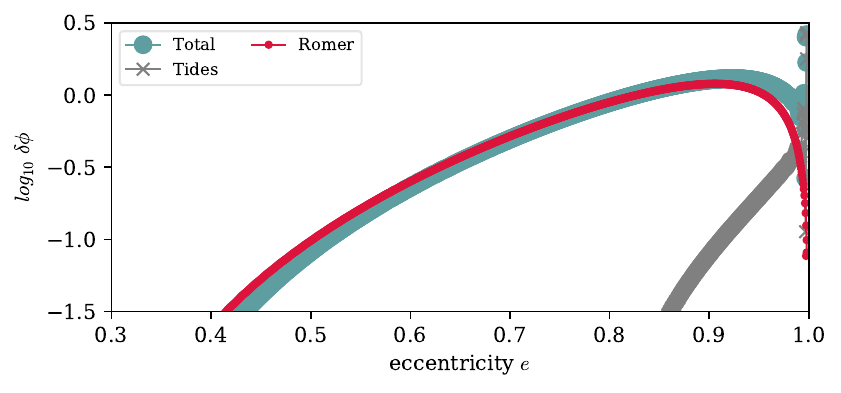}
\caption{{\bf Chaotic 3-body Scattering Merger.}
Observable differences between the 3-body-perturbed binary and
the 2-body-reference binary shown in Fig. \ref{fig:GW3bodyorbit_ill}, and discussed in Sec. \ref{sec:Example and Numerical Methods}.
We see how the unique features
we derived in Sec. \ref{sec:Analytical Models}, such as a maximum in the GW phase shift from Rømer delay, are still present here in the case
of a chaotic interaction. Larger differences between the outer-orbit being eccentric compared to circular, will be explored in upcoming papers.
\label{fig:GW3bodyorbit_PHASEetc} }
\end{figure}

One of the most reliable mechanisms for producing BBH mergers dynamically is through 3-body scatterings in dense stellar systems,
in particular GCs \citep{2018PhRvD..97j3014S, 2018PhRvL.120o1101R, 2018ApJ...855..124S, 2019ApJ...871...91Z, 2020MNRAS.492.2936A}.
In this scenario the heavier BHs in the cluster sink to the center, also known as mass segregation, to form a small sub-cluster of black holes with
relatively high density. Here the BHs pair up to form soft and wide binaries, that subsequently undergo interactions with the single BH population,
also known as 3-body interactions. These interactions not only stabilise the cluster through dynamical heating, but also bring the BBHs
close enough together for them to occasionally undergo a GW merger either inside or outside of the cluster \citep[e.g.][]{JSDJ18}.
In \cite{2018PhRvD..97j3014S}, it was in particular shown that about $10\%$ of all BBH mergers formed dynamically in GCs will merge inside the cluster
while being bound to the single BH that triggered the merger, often referred to as a chaotic `3-body merger'.

Fig. \ref{fig:GW3bodyorbit_ill} shows an example of the formation of such a 3-body merger, where the three panels in Fig. \ref{fig:3dviews}
show the same example from three different observer positions in {\it Stereoscopic Vision} (see caption for more details).
The initial binary is composed of two BHs each with a mass
of $15\ M_{\odot}$ at a separation of $a_0 = 0.01\ AU$, where the incoming BH has a mass of $5M_{\odot}$.
On the figures are also overplotted a
2-body-reference binary, as we did in the more idealised cases shown above. Fig. \ref{fig:GW3bodyorbit_PHASEetc} shows the
corresponding differences between the 3-body-perturbed binary and our 2-body-reference binary.
As seen for this particular system, the maximum GW phase shift is close to $\sim 1$ radian. Note also in Fig. \ref{fig:GW3bodyorbit_ill}, that the
BBH as it spirals in, actually evolves close to one full orbit around the bound single BH. This gives rise to additional modulation
effects than the one we worked out in Sec. \ref{sec:Analytical Models} above, but they are of course naturally included in
our numerical results shown here.
However, the main characteristics of this chaotically assembled eccentric GW phase shift source are the same as in our considered idealized analytical case;
One sees a clear peak in the GW phase shift around $e \sim 0.95$, and a corresponding rapid decrease as a function of decreasing eccentricity $e$.
An effect that also remains to be quantified, is the effect from the outer orbit to be eccentric, as this effectively will give rise to a varying
curvature or effective radius, $R$, along the orbit. It can here happen, e.g., that the BBH inspiral passes
through pericenter of the outer orbit, which could give rise to an additional increase in the GW phase shift off-centred from the maximum
$e\sim0.95$ that exists in the circular outer-orbit case. Other effects can be gravitational redshift, which could dominate over the Rømer delay depending on the
outer orbit eccentricity and position of the observer. We will quantify these effects in upcoming papers, and link the
overall GW phase shift distribution at different GW frequencies to the specific underlying dynamical channels and properties of the environment.

\section{Conclusions}

In this paper, we have quantified the GW phase shift appearing in the GW-form of eccentric inspirals evolving near a perturber, as a result of including
the time changing Doppler shift giving rise to a Rømer delay, and the tidal influence of the perturber.
By accurately measuring the time dependent GW phase shift, one can in principle infer the mass and orbits of the entire three-body system,
as in the case of binary pulsars, which offer unique possibilities for learning about the formation environment- and mechanisms of
dynamically formed BBH mergers \citep[see also][]{2017ApJ...834..200M}.
With upcoming GW detectors such as ET and CE that are expected to achieve SNRs of $10^{2}-10^{3}$ for BBH mergers,
such GW phase shift measurements will become one of the most powerful ways of probing the origin of individual GW sources using GWs alone.

Previous work has mainly focused on 
circular binaries on circular orbits near heavy perturbers; however, we point out that 
the most likely case is that the majority of sources with nearby perturbers must have been eccentric in the past, while some might even be at the time of observation.
The question of how the GW phase shift changes as a function of time, eccentricity, and GW frequency is therefore crucial to understand for making realistic
couplings between theory and observations. 

For our work we have developed a new numerical approach that is based on a novel technique, where we construct an unperturbed
reference binary by evolving the $\PN$ equations backwards
from a point near merger. The properties of this 2-body-reference binary is compared to the real 3-body-perturbed binary,
from which we can quantify, for the first time, the tiniest variations in, e.g., the observed GW signal emitted by the observed, 3-body-perturbed binary.
In our framework so far, we only include dynamical perturbations, and the contributions from Rømer delay, as these effects were argued in \cite{2017ApJ...834..200M} to 
likely be the dominating for most astrophysical systems. 
We do however plan on including more relativistic effects in upcoming versions.

We further present an analytical solution, that is an extension to the well-known GW phase shift relations in the circular case,
but now with the inclusion of
eccentricity, $e$. This gives, in particular, rise to a new and unique relation between maximum GW phase shift, $\Delta{\phi}$, and eccentricity, $e$.
Including eccentricity is crucial, as the most likely GW phase shift candidates will have formed dynamically, or at least in a
dynamical environment, which means their initial eccentricity at assembly will be non-zero.

Besides a few idealized examples of an eccentric, merging binary on a circular orbit around a central BH, 
we have for the first time also considered GW phase shifts from chaotic 3-body scatterings, that will occur frequently in GCs and
other dense stellar systems. The chaotic nature of this channel and corresponding GW phase shifts call for a more statistical approach
to make predictions for different systems, which we plan on doing in upcoming papers.

From this first study of the observable perturbed GW signal from an eccentric BBH inspiraling on an orbit around a third perturbing BH we
identify the following main results:

\begin{enumerate}

\item{
The Rømer delay, which is linked to the non-relativistic Doppler effect, leads to a GW phase shift $\Delta{\phi}$ of the eccentric BBH orbiting the third BH, which we find evolves
in a unique way as a function of eccentricity, $e$, as $\Delta{\phi} \propto 2e^{78/19}(1-e^2)^{1/2}g(e)^{13/2}$ (see Eq. \ref{eq:dphi_e_e1lim}).
This form does not depend on either the BH masses or any length scales; it only depends on eccentricity, which
implies that the maximum GW phase shift, and how it evolves near its maximum value, can be deduced from this simple equation.
For example, one can see that the GW phase shift is maximized when the BBH reaches an eccentricity of $e \sim 0.95$,
and already reduced by a factor of $10$ when $e\sim 0.5$ (see Sec. \ref{sec:Maximum GW Phase Shift}).
This strongly suggests that for GW phase shifts to be observable for dynamically assembled sources, we need to catch them in their
eccentric phase. Note here that these are general statements, where details
will depend trivially on the position of the observer.}

\item{
The tidal influence from the third BH can give rise to significant changes to the orbital parameters of the inspiraling BBH right after assembly,
which first results in clear wave-like perturbations to the GW peak frequency signal (Fig. \ref{fig:Ex1_dt_dphi_etc}).
The wave period is directly related to the $1\PN+2\PN$-precession period of the inspiraling BBH, which, if observed, will put tighter constrains
on the system.
The tidal coupling also affects the later stages of the inspiraling BBH, as the maximum GW phase shift effectively depends on the initial conditions, which sets
the merger time and thereby essentially how much the binary COM motion will differ from a standard reference binary evolving on a
straight line (Sec. \ref{sec:General Relations}). As a result, we argue and find that when tidal interactions are taken into account, the BBH orbital
parameters can in some cases be changed to result in much larger GW phase shifts than the analytical models predict.
}

\item{
The GW peak frequency of the BBH when it reaches its maximum GW phase shift is very close to the GW peak frequency at
formation (Sec. \ref{eq:fmf0_e}), i.e. when $f_m \sim f_0$.
This implies that sources that are expected to give rise to the largest observable GW phase shifts (observable in terms of peak frequency)
also will have to form with $f_0$ near the observable band.
Many channels involving three-body systems are expected to give rise to at least a fraction of sources that form with $f_0$ directly
in the observable band \cite[e.g.][]{2022Natur.603..237S}. Eccentric sources with GW phase shifts are therefore expected to exists, the question is if the third object is close enough for the shift to be large enough to be observable. That depends not only on the chaotic nature of the systems, but also more broadly on the velocity dispersion of the underlying environment, as this sets the characteristic size of the three-body system and thereby roughly
the magnitude of the GW phase shift.
}

\item{
The GW phase shift as a function of GW peak frequency, $f_p$, is of direct interest in terms of observations, and has been shown in the circular case to scale
$\propto f^{-13/3}$. However, when the BBH is evolving with a non-zero eccentricity, this simple relation does not hold, instead we find that an
approximate solution that takes the form $\propto  f^{-13/3} \times (1+({f_0}/{f}))^7(1-({f_0}/{f}))^{1/2}$. This implies that the GW phase shift
actually increases faster towards lower GW peak frequencies compared to the 
circular case, with a peak that can be $\sim 30$ times larger near the binary formation frequency
compared to the case of circular binaries. This we confirmed with our simulations shown in Fig. \ref{fig:dphi_freq}, and as further discussed in Sec. \ref{sec:Evolution with Frequency}.
}

\end{enumerate}

With the theory and new numerical framework presented in this paper, we will in upcoming papers study the outcomes from a large ensemble of few-body scatterings taking place in astrophysical systems from GCs to AGN disks, and statistically quantify the observable possibilities in relation to LISA, DECIGO, 3G-detectors, and LVK.

\section{Acknowledgments}

It is a pleasure to thank Martin Pessah, Zihan Zhou, and Yan Yu for enlightening discussions.
The authors also thank the Research Center for the Early Universe (RESCEU) at Tokyo University
where part of this work was completed.
J.S., K.H. and L.Z. are supported by the Villum Fonden
grant No. 29466, and by the ERC Starting
Grant no. 101043143 — BlackHoleMergs.
B.L. gratefully acknowledges support from the European Union’s
Horizon 2021 research and innovation programme under the Marie Sklodowska-Curie
grant agreement No.
101065374. 

\newpage

\appendix

Below we show a few specific cases to illustrate the importance of using numerical prescriptions, such as the $\PN$ framework we have introduced that
mixes forward- and backwards evolutions (real- and reference binary, respectively), for solving the binary and single problem with corresponding GW
phase shifts. We highlight below in particular the sensitivity to the initial conditions, and the importance of including the 1$\PN$, 2$\PN$ energy
conserving precession terms.

\section{Sensitivity to Initial Conditions and Tides:}\label{app:Sensitivity to Initial Conditions and Tides:}

The three-body systems that are expected to give rise to the largest GW phase shifts, are generally also the ones where the binary is assembled with a
SMA about the size of the Hill sphere (see Sec. \ref{sec:Maximum GW Phase Shift} and Fig. \ref{fig:BS_ill}).
The reason is simply that the less compact the binary is, the larger the merger time generally will be, and the
more the binary COM trajectory from assembly to merger will differ from a straight line. The most relevant three-body systems are therefore
the ones where the initial BBH is most easily perturbed by tidal interactions during its early evolutionary stages of the inspiral phase.
We here show the sensitivity to the initial conditions when the BBH is assembled near the Hill sphere, and how this impact the evolution
and corresponding GW phase shift.

\begin{figure*}[h]
\centering
\subfloat{
\includegraphics[width=0.5\columnwidth]{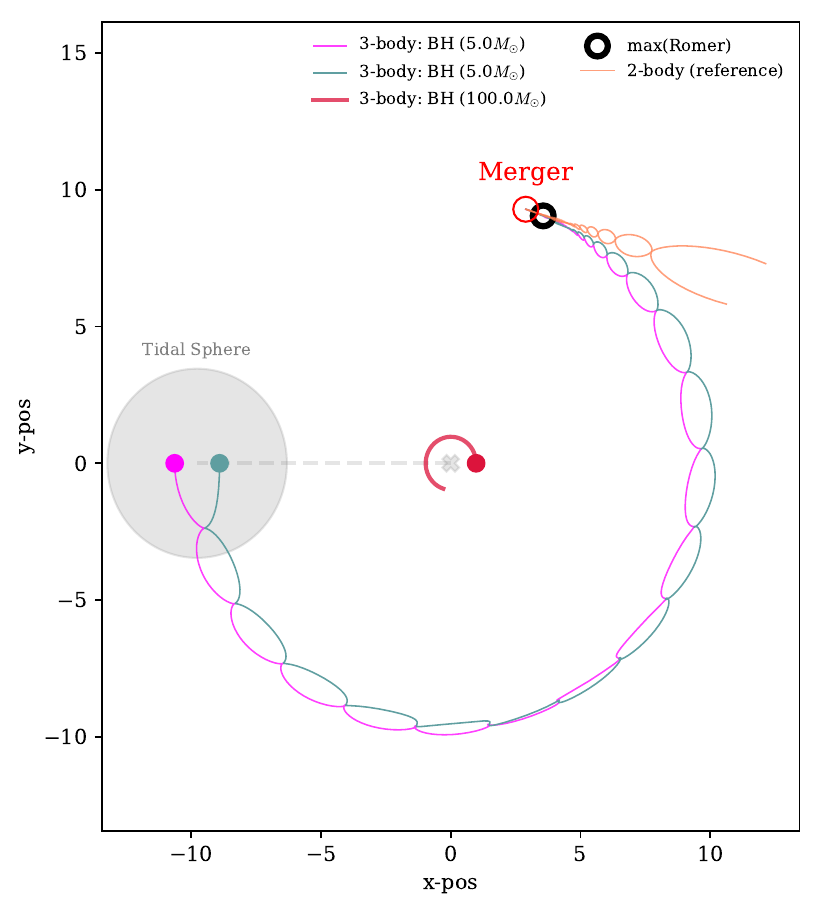}
\includegraphics[width=0.5\columnwidth]{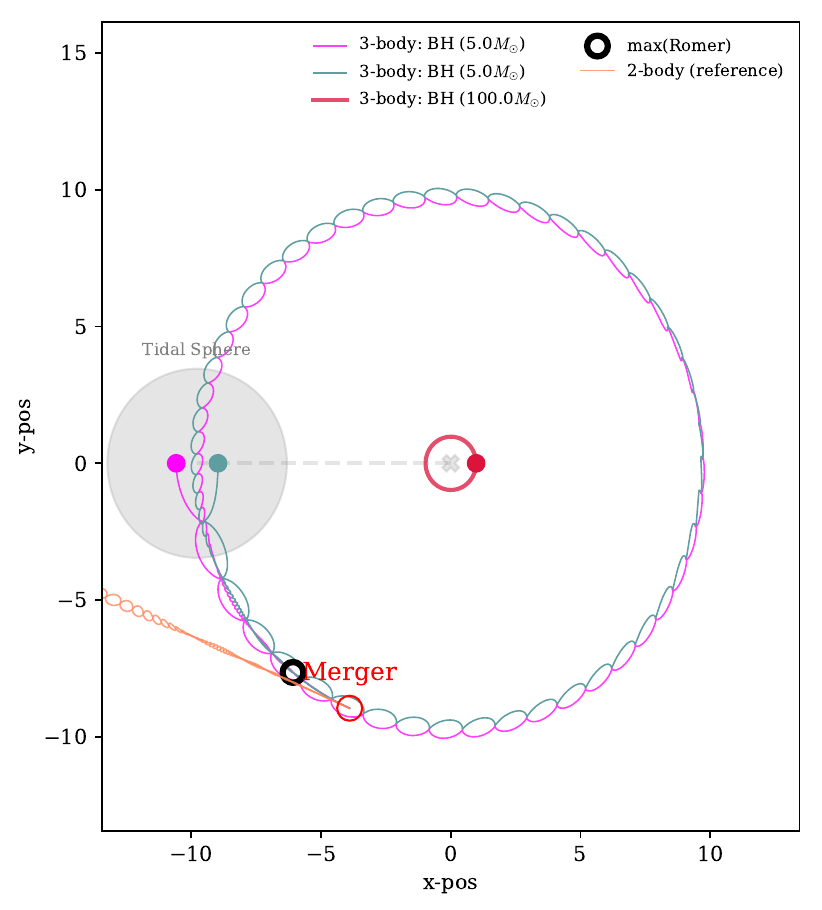}}
\qquad
\subfloat{
\includegraphics[width=0.5\columnwidth]{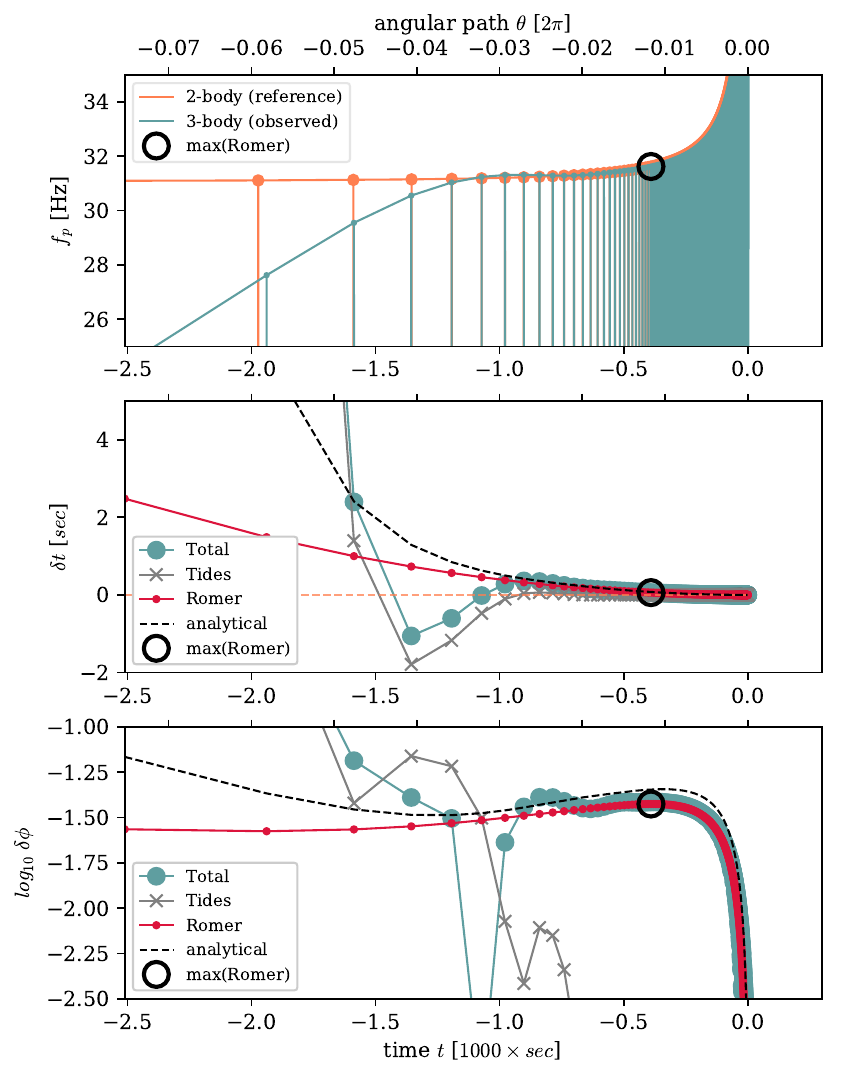}
\includegraphics[width=0.5\columnwidth]{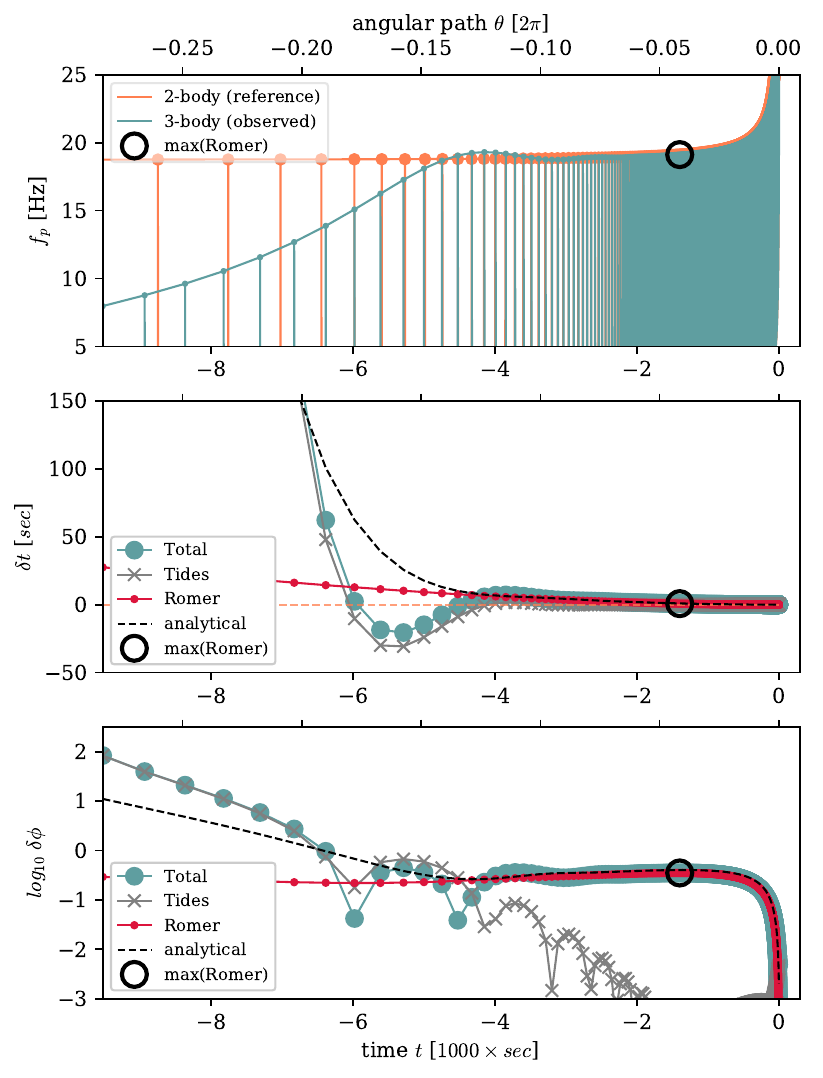}}
\caption{{\bf Sensitivity to Initial Conditions and Tides:} Examples similar to the one introduced in Fig. \ref{fig:Ex1_orbits}, differing only slightly in the
initial SMA $a_0$ as seen in the figure. The significant differences in both orbital evolution and corresponding GW phase shifts between the left and right panel,
are due to tidal couplings between the binary and the central BH early times in their inspiral. This coupling not only depends on the binary SMA and eccentricity, $a,e$,
but also on the orbital orientation and binary phase. Numerical studies, as the one we show in this paper, are therefore crucial in making accurate
predictions for how to link theory to observations.}
\label{fig:APP1}
\end{figure*}

Fig. \ref{fig:APP1} shows essentially the same setup as the one studied in Sec. \ref{sec:Example and Numerical Methods}, but with minor changes to
the initial SMA. As seen, in the left panel the tidal interaction between the binary and the third BH leads to an irregular inspiral at early times, after
which the tidal forces perturbs the BBH orbit into a combination of values the lead to a fast merger. The short merger time implies a relative high
GW peak frequency, as seen to be around $30$ Hz at formation, but also a corresponding small GW phase shift as seen in the bottom panel and
expected from e.g. Eq. \ref{eq:dphi_e_e1lim}, which shows $\Delta{\phi} \propto f_0^{-13/3}$.
In the right panel of Fig. \ref{fig:APP1}, the tides lead instead to a more longterm and gradual inspiral, which evolves over the entire outer orbit.
The final assembly at inspiral takes place with a GW peak frequency of around $20$ Hz, which therefore leads to a GW phase shift that is
$1-2$ orders-of-magnitude larger than the evolution shown in the left panel. This example clearly illustrates the chaotic nature of the tidal-
and $\PN$ couplings at early times, and how they affect the late time evolution and corresponding maximum GW phase shift.
The effect from tides therefore brings both opportunities and difficulties that we will explore in upcoming papers.   

\section{Importance of the 1$\PN$, 2$\PN$ precession terms:}\label{sec:app Importance of the 12PN}

The dissipative 2.5$\PN$ term, is the one that leads to the gradual inspiral and eventual merger of the BBH. However,
as we have argued in the above Sec. \ref{app:Sensitivity to Initial Conditions and Tides:} and
in Sec. \ref{sec:Example and Numerical Methods}, the BBHs that are most likely to give rise to measureble GW shifts are also the ones that
are most likely to be in tidal contact with the third BH during the early stages of their inspiral phase. This tidal influence on the binary
was in \cite{2023PhRvD.107l2001R} argued, using an analytical `burst timing model', to lead to a gradual build up in the GW phase shift, which could be used for
probing the presence of the third BH; however, the burst timing model that was developed for this, did not include the conservative
precession terms 1$\PN$, 2$\PN$. We here show the crucial importance of including these terms, as we also argued in
Sec. \ref{sec:Numerical Setup and Observers}.

\begin{figure*}[h]
\centering
\subfloat{
\includegraphics[width=0.5\columnwidth]{SET1_W12_EX4txt_ijk_orbits_xyz.pdf}
\includegraphics[width=0.5\columnwidth]{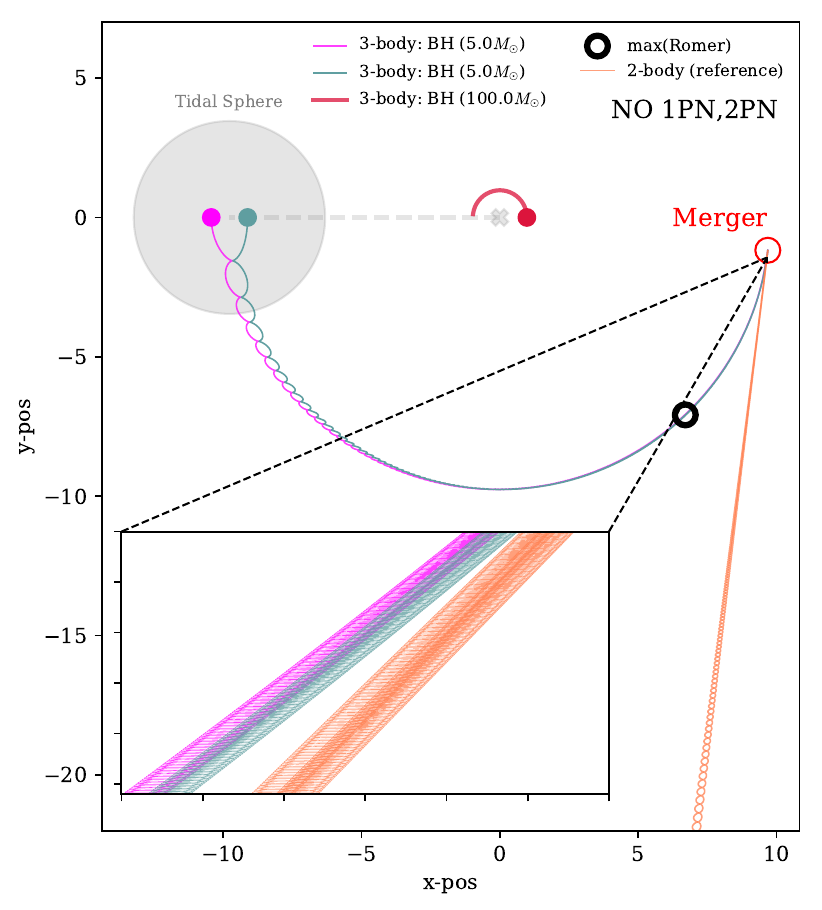}}
\qquad
\subfloat{
\includegraphics[width=0.5\columnwidth]{SET1_W12_EX4txt_dt_peak_analysis_1.pdf}
\includegraphics[width=0.5\columnwidth]{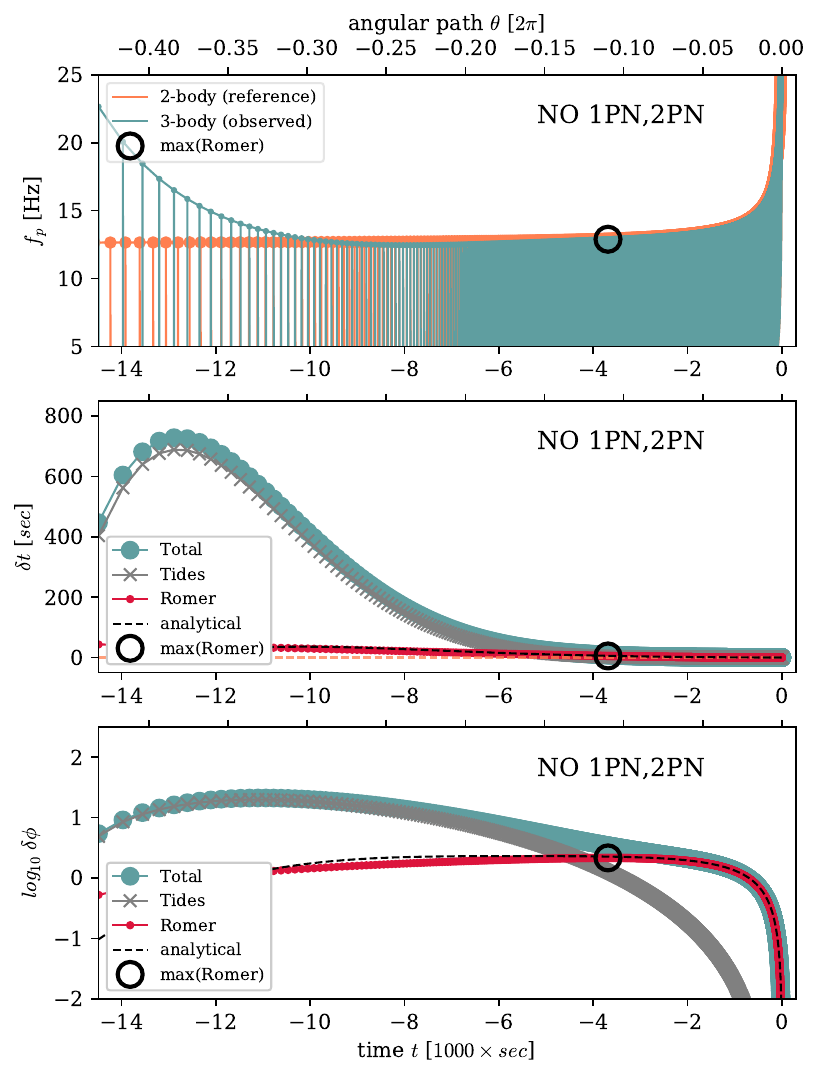}}
\caption{{\bf Importance of the 1$\PN$ ,2$\PN$ precession terms:} Left panel shows our considered example from Fig. \ref{fig:Ex1_orbits}, where the right panel
shows a similar system for comparison, but without the inclusion of the 1$\PN$, 2$\PN$ precession terms in the EOM (`NO 1PN,2PN').
As seen, when $\PN$ precession is not included, tides from the central BH3 can without any problem gradually build up to result in large
time- and GW phase shifts. This is incorrect, and serves therefore as a clear illustration to why the conservative 1$\PN$, 2$\PN$ terms are crucial.}
\label{fig:APP2}
\end{figure*}

We now consider Fig. \ref{fig:APP2}, which shows in the left panel the same example as we considered in Sec. \ref{sec:Example and Numerical Methods},
and in the right panel results from similar initial conditions, but without the inclusion of the conservative 1$\PN$, 2$\PN$ terms. As seen in the right panel,
when the BBH by construction is not able to precess, the tidal force on the BBH from the third BH is simply gradually building up to result in a
predictable and smooth evolution of the GW phase shift, that as a result, is driven to a huge value at its peak. This is clearly an unphysical result,
which illustrates the importance of a correct dynamical implementation for evolving such systems.
The conservative 1$\PN$, 2$\PN$ terms have also been shown to play a major role in the evolution of hierarchical three-body systems \citep[e.g.][]{2016ARA&A..54..441N}.

The fact that the few-body evolution and corresponding GW phase shift depend so strongly on how the binary tidally couples to the third object
also brings some concerns to our model, as this also is not fully accurate mainly due to approximations in the $\PN$ formalism.
However, while more accurate models can be adapted, we do believe that we have been able to resolve the main effects of the problem
using our model. Further effects will be explored in upcoming papers, such as outer orbit eccentricity and binary tilt, as well as black hole spins;
effects that all will result in their own GW phase shift contribution.

\newpage

\bibliographystyle{aasjournal}
\bibliography{NbodyTides_papers}

\end{document}